%% file: latproc.tex
\documentclass{PoS}
\usepackage{amsmath,amssymb,graphicx,slashed}
\graphicspath{{figs/}}

\title{Looking for $\mathbf{U(1)_A}$ Restoration in Hot QCD with Domain Wall Fermions}

\ShortTitle{Looking for $U(1)_A$ Restoration}

\author{\speaker{Prasad Hegde}\\
        Brookhaven National Laboratory, Upton, NY 11973, USA.\\
        E-mail: \email{phegde@quark.phy.bnl.gov}}

\abstract{The effects of the axial anomaly are suppressed at high temperatures due to screening effects in the quark-gluon plasma. If the suppression is nearly complete close to the chiral transition temperature, this can have consequences for the nature of the phase transition. The use of a chiral action such as Domain Wall Fermions allows us to gain a deeper insight into the issue. Our lattice sizes were $16^3\times 8\times L_s$, with $L_s=32$ or 48, and our pion mass was approximately 200 MeV. We found that $\uonea$ stayed broken above the chiral transition. However the breaking was found to be due to topologically nontrivial configurations which raises the question as to whether it persists in the thermodynamic limit. We also present results for the eigenvalue density of the Dirac operator. It is seen that although the density decreases dramatically across the chiral transition temperature, $\uonea$ still remains  broken at our current volume and quark mass due to the presence of zero modes.}

\FullConference{ The XXIX International Symposium on Lattice Field Theory - Lattice 2011\\
July 10-16, 2011\\
Squaw Valley, Lake Tahoe, California}

\newcommand{\uonea}{U(1)_A}
\newcommand{\sunfa}{SU(2)_L\times SU(2)_R}

\newcommand{\ub}{\overline{u}}
\newcommand{\db}{\overline{d}}

\newcommand{\ul}{u_L}
\newcommand{\ur}{u_R}
\newcommand{\dl}{d_L}
\newcommand{\dr}{d_R}

\newcommand{\ubl}{\overline{u}_L}
\newcommand{\ubr}{\overline{u}_R}
\newcommand{\dbl}{\overline{d}_L}
\newcommand{\dbr}{\overline{d}_R}

\newcommand{\psib}   {\overline{\psi}}

\newcommand{\pbp}    {\psib\psi}
\newcommand{\qtop}   {Q_\text{top}}
\newcommand{\chit}   {\chi_\text{top}}
\newcommand{\pbgp}   {\psib\gamma_5\psi}
\newcommand{\lmin}   {\lambda_\text{min}}
\newcommand{\lpbpr}  {\langle\psib\psi\rangle}
\newcommand{\lpbgpr} {\langle\psib\gamma_5\psi\rangle}

\newcommand{\density}{\rho(\lambda,m)}

\newcommand{\Ns}{N_\sigma}

\newcommand{\I}   {\mathrm{i}}
\newcommand{\E}   {\mathrm{e}}
\newcommand{\df}  {\mathrm{d}}

\newcommand{\mres}{m_\text{res}}

\newcommand{\chic} {\chi_\delta}
\newcommand{\chid} {\chi_\text{disc}}
\newcommand{\chigc}{\chi_\pi}
\newcommand{\chigd}{\chi_\text{5,disc}}
\newcommand{\pmd}{\chi_\pi-\chi_\delta}

\begin{document}
\section{Introduction}
\label{sec:intro}
The Lagrangian of Quantum Chromodynamics (QCD) with $N_f$ massless flavors of quarks is invariant under a global $SU(N_f)_L \otimes SU(N_f)_R \otimes  U(1)_V \otimes U(1)_A$ symmetry. In the vacuum, the $SU(N_f)_L \otimes SU(N_f)_R$ chiral symmetry is spontaneously broken to a $SU(N_f)_V$ subgroup, corresponding to flavor symmetry. This spontaneous breaking of chiral symmetry gives rise to a nonvanishing expectation value $\lpbpr$ of the chiral condensate.

The axial $U(1)_A$ symmetry of the QCD Lagrangian on the other hand is broken by the axial anomaly. The inclusion of quantum fluctuations leads, at the perturbative level itself, to non-conservation of the axial current; this is the famous Adler-Bell-Jackiw anomaly \cite{Adler:1969gk,Bell:1969ts}:
\begin{equation}
\big\langle \partial_\mu j^{\mu 5} \big\rangle = -\frac{\alpha_s}{4\pi}\big\langle\epsilon^{\alpha\beta\gamma\delta}
F^a_{\alpha\beta}F^a_{\gamma\delta}\big\rangle .
\label{eq:anomaly} 
\end{equation}

In QCD, the anomaly implies global non-conservation of axial charge. Naively, integrating Eq.~\eqref{eq:anomaly} over all spacetime should give zero since the left-hand side is a total divergence. However there exist special gauge field configurations in QCD for which the integral of the right-hand side is not zero. These are the configurations with nontrivial topology~\cite{'tHooft:1976up}. All such configurations must be included in the path-integral. Anomalous contributions arise for any observable for which the contribution from such configurations is unsuppressed.

\subsection{Effective $\mathbf{\uonea}$ Restoration}
\label{ssec:effective}
A common example of a phase transition in several finite-temperature field theories is the restoration of a spontaneously broken global symmetry. This is the case with chiral symmetry in QCD as well. For $N_f=2$ in the massless limit, the phase transition is expected to be second-order and belonging to the $O(4)$ universality class. When the quarks are massive, this transition becomes a crossover.

By contrast axial symmetry is broken at the perturbative level itself. There is thus no question of its complete restoration at any temperature. However as we have already seen, anomaly-related effects arise from the existence of topologically nontrivial configurations. The action for these configurations is proportional to $\alpha_s^{-1}$. Such actions are therefore Boltzmann-suppressed due to the screening of the coupling constant at high temperatures~\cite{Gross:1980br}. Although there is always some amount of $\uonea$ breaking below $T=\infty$, it is conceivable that this suppression is nearly complete by some temperature that is not too high. We may then speak of an \emph{effective} restoration of the axial symmetry.

If this temperature is close to the chiral phase transition temperature $T_c$, then the effective restoration of $\uonea$ can have interesting phenomenological consequences. The standard picture of a second-order phase transition is based on the assumption that $\uonea$ breaking is substantial near $T_c$.\footnote{Note that the chiral condensate $\lpbpr$, which signals chiral symmetry breaking, also breaks $\uonea$. Consequently there is no question of $\uonea$ being restored before $\sunfa$ is.} If this is not the case, then the phase transition may even be first order~\cite{Pisarski:1983ms}. Understanding the contribution of $\uonea$ is thus essential to mapping the phase diagram of QCD.

\section{Domain-Wall Fermions and DSDR}
\label{sec:dsdr}
Chiral symmetry restoration and effective $\uonea$ restoration at high temperatures are both non-perturbative phenomena whose reliable study demands the use of nonperturbative techniques. Currently, lattice QCD is certainly the most viable and reliable such technique. Extensive lattice QCD studies of chiral symmetry restoration have already been carried out (for a review and summary see~\cite{DeTar:2009ef,Mukherjee:2011td}). The question of $\uonea$ restoration too has been investigated before~\cite{Bernard:1996iz,Chandrasekharan:1998yx,Kogut:1998rh,Cheng:2010fe}.

However such studies have almost always been carried out with staggered fermions. For these fermions the issues of chiral symmetry, the anomaly and the relation between the anomaly and the index theorem are very subtle~\cite{Sharpe:2006re, Donald:2011if, Adams:2009eb}. Hence further studies using different fermion discretization schemes are certainly welcome.

Domain Wall Fermions are a fermion discretization scheme that preserves the full $SU(N_f)_L\times SU(N_f)_R$ chiral symmetry of continuum QCD and also reproduces the correct anomaly even at nonzero values of the lattice spacing~\cite{Kaplan:1992bt}. The domain wall formulation is one of five-dimensional fermions whose low-energy spectrum is four-dimensional and also, when the fifth dimension is infinite in extent, exactly chiral. The gauge fields remain four-dimensional and couple to the fermions in the usual way. For finite fifth dimension, the residual chiral symmetry breaking manifests itself at low energies as an additive shift $\mres$ of the bare quark mass~\cite{Kaplan:2009yg}.

The QCD phase transition has been studied before with domain wall fermions~\cite{Chen:2000zu,Cheng:2009be}. A challenge encountered in the most recent study was the rapid variation of $\mres$ as one moved toward stronger coupling which made it difficult to keep the pion mass fixed throughout the temperature range studied~\cite{Cheng:2009be}. The use of improved gauge actions such as the Iwasaki action results in a smaller value of $\mres$ overall but cannot arrest the rapid growth of $\mres$ as the temperature is decreased.

In an ongoing study of QCD thermodynamics using domain wall fermions by the HotQCD collaboration~\cite{Bazavov:2012aa}, $\mres$ was sought to be kept to a minimum through the use of the ``Dislocation Suppressing Determinant Ratio (DSDR).'' The usual Iwasaki gauge action was augmented with a ratio of Wilson determinants which suppressed the zero modes (dislocations) that contributed to $\mres$. To maintain adequate topological tunneling, the Wilson-Dirac mass was set equal to the domain-wall height $-M_0$ plus a small chirally twisted mass $\I\epsilon\gamma_5$ viz.~\cite{Vranas:2006zk,Fukaya:2006vs,Renfrew:2009wu}
\begin{equation}
\frac{\det\left[D_W^\dagger(-M_0 + \I \epsilon_f \gamma_5) D_W(-M_0 + \I \epsilon_f \gamma_5)\right]}
     {\det\left[D_W^\dagger(-M_0 + \I \epsilon_b \gamma_5) D_W(-M_0 + \I \epsilon_b \gamma_5)\right]}.
\label{eqn:DSDR}
\end{equation}

With this action we generated a few thousand configurations each at seven temperatures between 140 MeV and 200 MeV. Our lattice sizes were $16^3\times 8\times L_s$ with $L_s=32$ for $T\geqslant 160$ MeV and $L_s=48$ at lower temperatures. Before generating these configurations, we generated several zero-temperature ensembles at several values of the coupling $\beta$ both to set the scale and to determine the residual mass $\mres$. The input light and strange quark masses were then chosen so as to keep the kaon physical and the pion mass fixed at 200 MeV for all $\beta$; this defined our Line of Constant Physics.

At each temperature, we measured $\lpbpr$, $\lpbgpr$ and the corresponding disconnected susceptibilities. We also measured the flavored scalar $(\delta)$, pseudoscalar $(\pi)$, vector $(\rho)$ and axial vector $(a_1)$ correlators to be discussed below. Separately, we also measured the topological charge $\qtop$ for each configuration through the use of cooling and smeared gauge field operators~\cite{hep-lat/0612005}. From this we calculated $\langle\qtop\rangle$, $\langle\vert\qtop\vert\rangle$ and the topological susceptibility $\chi_\text{top}$. Finally, we also measured the lowest hundred eigenvalues of the five-dimensional Dirac operator on each configuration in an effort to determine the eigenvalue density distribution $\rho(\lambda)$ (Section~\ref{sec:dirac_spectrum}). The physics behind the DSDR action, its performance and the results for the chiral phase transition have been presented by M.~Cheng at this conference~\cite{Cheng:2011lt}. A complete description of our ensembles, scale determination and measurements is also forthcoming~\cite{Bazavov:2012aa}.

\begin{figure}[!tbh]
\centering
\vspace{0.025\textheight}
\includegraphics[width=0.5\textwidth]{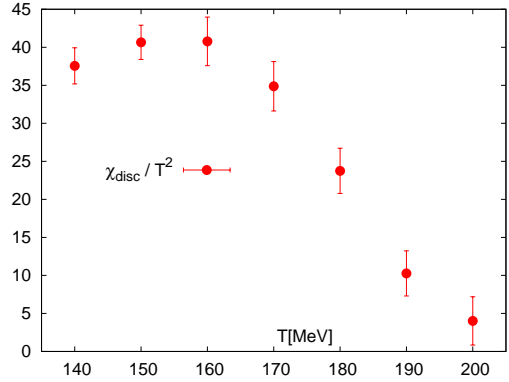}
\caption{The one-flavor disconnected chiral susceptibility for the light quark. The transition region is broad, as might be expected of a crossover, with a peak near $T\approx 160$ MeV.\label{fig:Tc}}
\end{figure}

Fig.~\ref{fig:Tc} plots the disconnected chiral susceptibility as a function of the temperature. The susceptibility peaks at 160 MeV; accordingly we take that to be the approximate value of the chiral phase transition temperature $T_c$. Since the phase transition is expected to merely be a crossover for $m_\pi>0$, this value only serves as a reference when discussing the possibility of $\uonea$ restoration.

\section{Symmetries, Correlators and Susceptibilities}
\label{sec:symmetries}
\begin{figure}[!tb]
\centering
\includegraphics[width=0.6\textwidth]{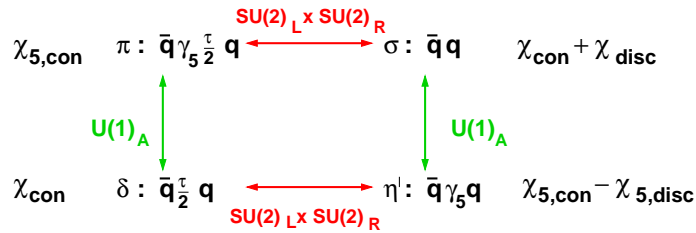}
\caption{The symmetries $SU(N_f)_L\times SU(N_f)_R$ and $\uonea$ relate mesons in different spin-flavor channels. The above diagram summarizes these relations for $N_f=2$.}
\label{fig:relations}
\end{figure}

The influence of a symmetry is seen on the appropriate correlators of Dirac bilinears. In the scalar-pseudoscalar sector, for $N_f=2$, we have two iso-triplet correlators:
\begin{subequations}
\begin{align}
C_\delta(x) &= \big\langle  \ub d(x)\;\db u(0) \big\rangle,\\
C_\pi(x)    &= \big\langle\I\ub\gamma_5 d(x)\;\I\db\gamma_5 u(0) \big\rangle,
\end{align}
\label{eq:pi_delta_corr}
\end{subequations}
as well as two iso-singlet correlators:
\begin{subequations}
\begin{align}
C_\sigma(x) &= \Big\langle \left(\ub u(x)+\db d(x)\right)\; \left(\ub u(0)+\db d(0)\right) \Big\rangle,\\
C_{\eta'}(x)&= \Big\langle \left(\I\ub\gamma_5 u(x)+\I\db\gamma_5 d(x)\right)\;
                           \left(\I\ub\gamma_5 u(0)+\I\db\gamma_5 d(0)\right) \Big\rangle.
\end{align}
\label{eq:sigma_etapr_corr}
\end{subequations}
The $\delta$ and $\pi$ correlators receive contributions only from diagrams with connected quark lines and are thus easier to measure. The $\sigma$ and the $\eta'$ on the other hand receive contributions from diagrams with connected as well as disconnected quark lines. The connected parts of these correlators are just the $\delta$ and the $\pi$ respectively. The full correlator however is obtained only after this part is canceled by a similar contribution from the disconnected piece. This makes the $\sigma$ and $\eta'$ correlators much harder to measure.

These correlators transform into one another under a chiral or an axial rotation, as summarized in Fig.~\ref{fig:relations}. This implies for e.g. that the $\pi$ and the $\sigma$ ($\delta$) correlators become identical when chiral (axial) symmetry is restored.

In addition to the correlators in eq.~\eqref{eq:pi_delta_corr} we also calculated the connected vector and axial vector correlators viz.
\begin{align}
&&
C_\rho (x) &= \big\langle  \ub\gamma_\mu d(x)\;\db \gamma_\mu u(0) \big\rangle, &
C_{a_1}(x)    &= \big\langle\I\ub\gamma_5 \gamma_\mu d(x)\;\I\db\gamma_5 \gamma_\mu u(0) \big\rangle.
&&
\label{eq:rho_a1_corr}
\end{align}

\begin{figure}[!tbh]
\centering
\hspace{-0.05\textwidth}%
\includegraphics[width=0.45\textwidth]{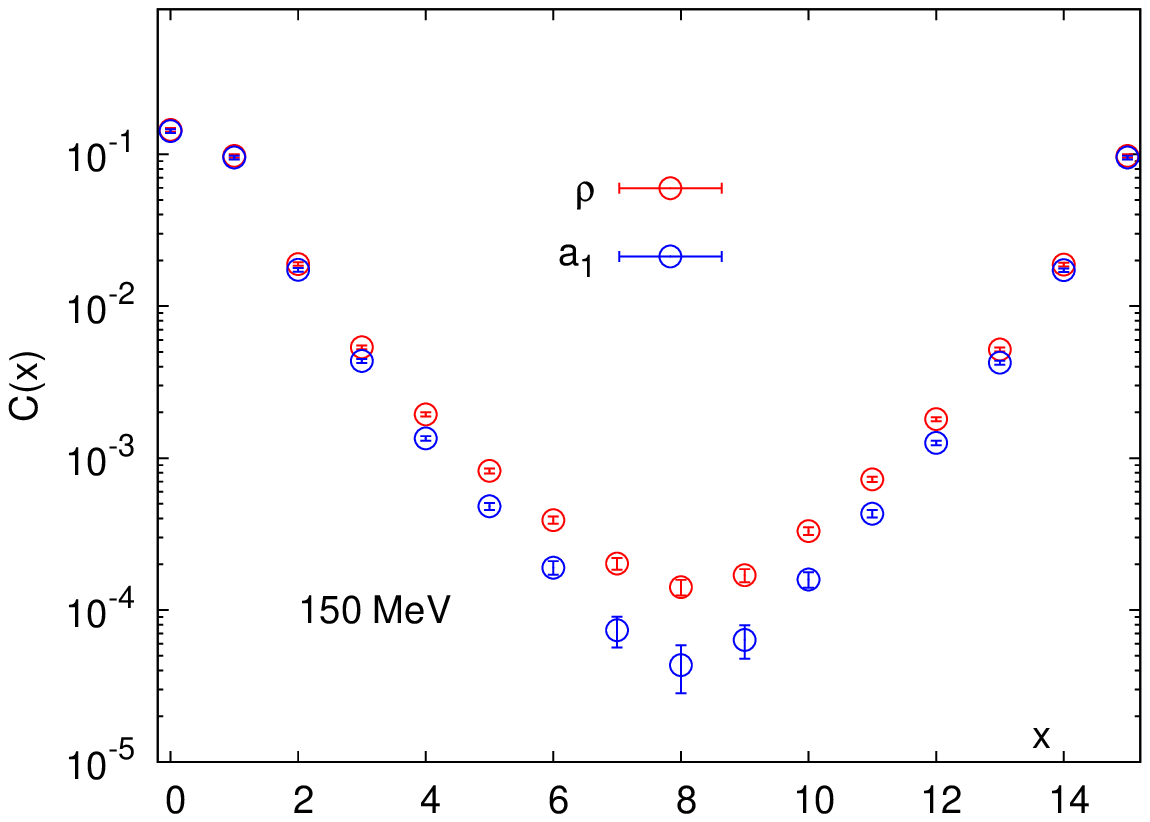}%
\hspace{0.05\textwidth}%
\includegraphics[width=0.45\textwidth]{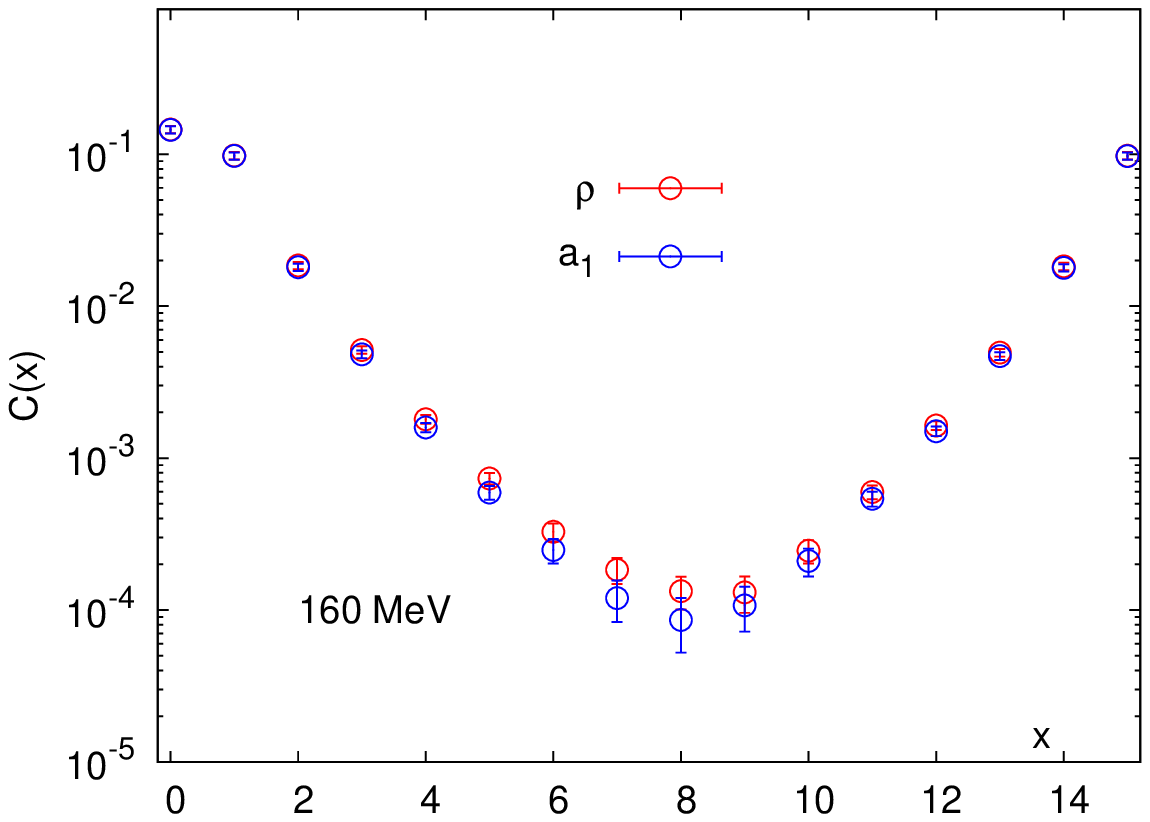}
\caption{The vector $(\rho)$ and axial vector $(a_1)$ correlators for $T=150$ MeV and 160 MeV respectively.}
\label{fig:vav_correlators}
\end{figure}

An axial rotation has no effect on the vector and axial vector correlators. The two are in fact related through chiral transformations and they become degenerate when chiral symmetry is restored. We plot these correlators at $T=150$ MeV and 160 MeV in Fig.~\ref{fig:vav_correlators}. Since the transition is a crossover the two only become exactly identical at very high temperatures, but they are nearly degenerate by $T=160$ MeV.

By integrating these correlators over the four-volume, we obtain the corresponding susceptibilities $\chi_\pi$, $\chi_\sigma$, etc. Just as for the correlators, one has connected and disconnected susceptibilities depending on the type of correlator being integrated. Furthermore, the disconnected parts of the $\sigma$ and $\eta'$ susceptibilities are equal to the disconnected susceptibilities $\chid$ and $\chigd$ viz.
\begin{align}
&& \chi_{\sigma,\text{disc}} = 
\big\langle\left(\pbp\right)^2\big\rangle -\big\langle\left(\pbp\right)\big\rangle^2 \equiv \chid && \text{and} && 
\chi_{\eta',\text{disc}}  = \big\langle\left(\pbgp\right)^2\big\rangle \equiv \chigd. &&
\label{eq:disc_susc}
\end{align}
The appropriate symmetry restoration gives rise to equalities among the different susceptibilities:
\begin{subequations}
\begin{align}
 && \chigc &= \chic + \chid && &\text{and} &&         \chic &= \chigc - \chigd. && \big[\sunfa\big] 
 \label{eq:susc-I} \\
 && \chigc &= \chic         && &\text{and} && \chic + \chid &= \chigc - \chigd. && \big[\uonea\big].
 \label{eq:susc-II}
\end{align}
\label{eq:susc_relations}
\end{subequations}
The difference $\chi_\pi-\chi_\delta$ must go to zero as $\uonea$ breaking is suppressed. Eq.~\eqref{eq:susc-I} tells us that this difference equals  $\chid$ once chiral symmetry is restored. Moreover, we see that chiral symmetry restoration implies that $\chid = \chigd$ whereas axial symmetry restoration implies the opposite, namely $\chid = -\chigd$. Either way, when \emph{both} chiral and axial symmetry are restored, one has $\chid=0=\chigd$. In other words, $\uonea$ restoration is signaled by a vanishing disconnected chiral susceptibility.


\begin{figure}[!tbh]
\centering
\includegraphics[width=0.5\textwidth]{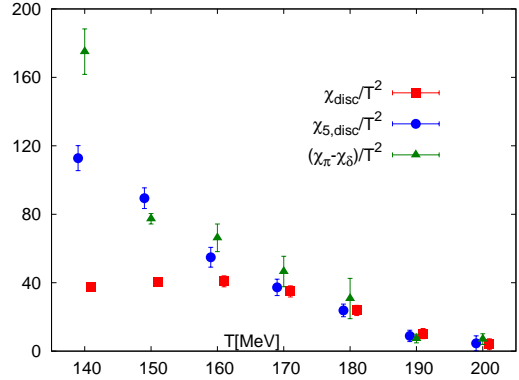}%
\caption{The susceptibilities $\chid$, $\chigd$ and $\pmd$ for each of the temperatures. All are very nearly equal from $T=170$ MeV onward. None of these susceptibilities vanishes for all the temperatures shown here. The red and blue points have been horizontally displaced by $\pm 1$ MeV for clarity.}
\label{fig:susceptibilities}
\end{figure}

Fig.~\ref{fig:susceptibilities} plots these susceptibilities for each of the temperatures that we studied. Although the equalities derived in Eqs.~\eqref{eq:susc_relations} are strictly valid only in the chiral limit, we see that $\chid$, $\chigd$ and $\pmd$ are almost equal to each other from about 170 MeV onwards. Furthermore, none of these susceptibilities is equal to zero even at $T=200$ MeV, the highest temperature that we studied. If we take $T_c\approx 160$ MeV, this would seem to suggest that $\uonea$ remains broken even at $T\approx1.25 T_c$.

\section{The Correlation with Topology}
\label{sec:topology}
\begin{figure}[!bht]
\includegraphics[width=0.45\textwidth]{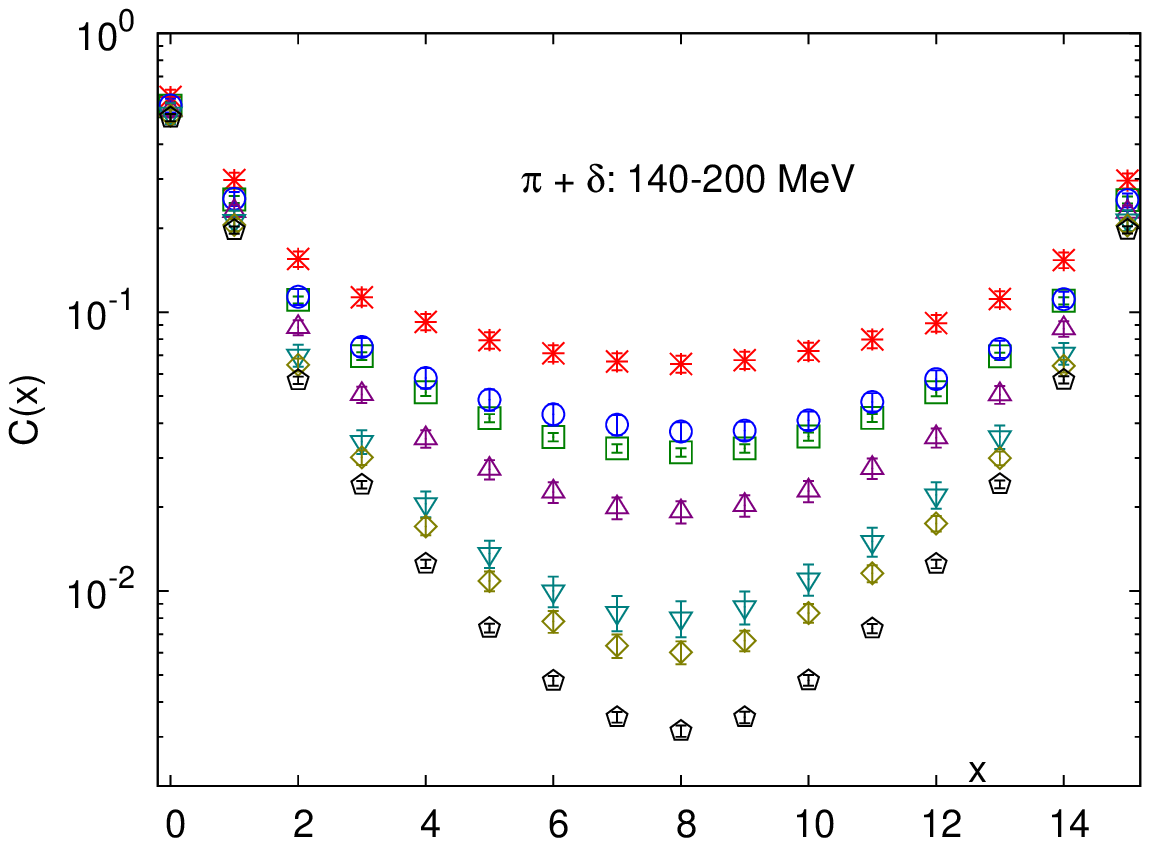}%
\hspace{0.05\textwidth}%
\includegraphics[width=0.45\textwidth]{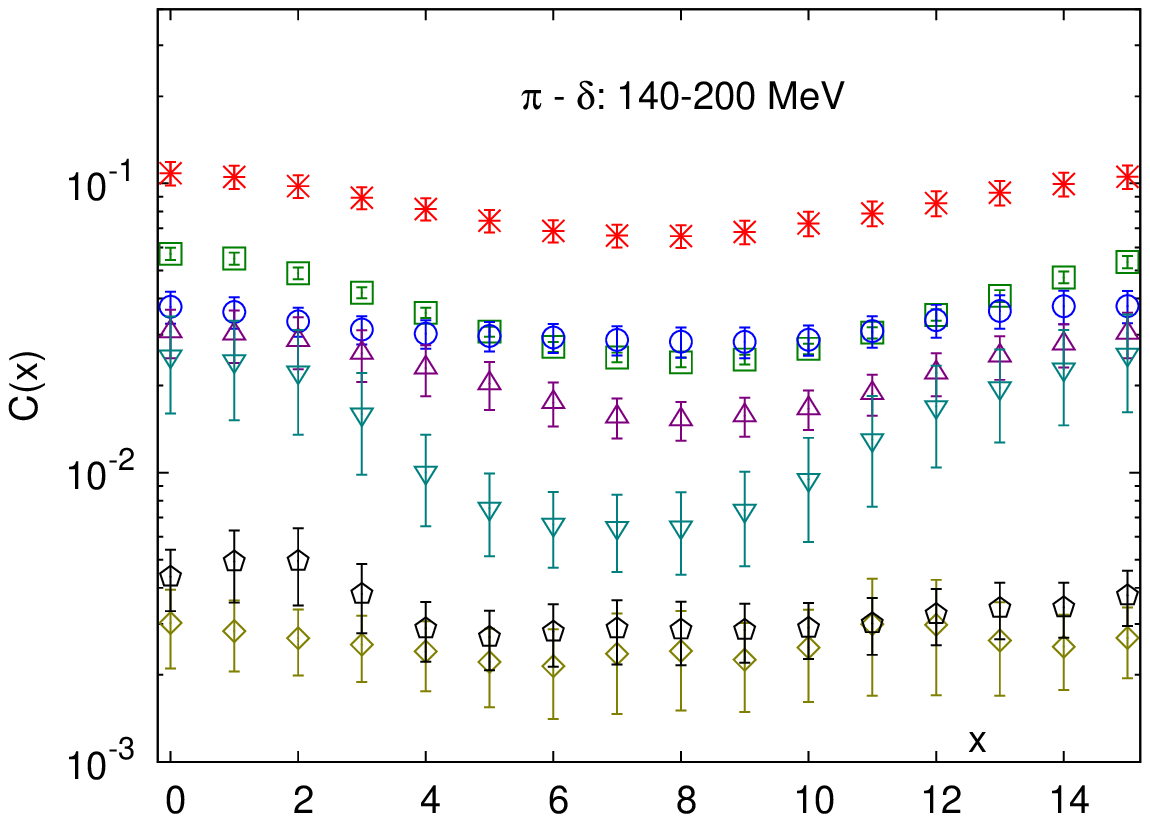}
\caption{(Left) The sum of the $\delta$ and $\pi$ correlators. The temperature increases from 140 to 200 MeV as one moves downward along the $y$-axis. (Right) The difference of the two correlators. The temperatures are identified by the same symbols as in the plot on the left.\label{fig:s_pm_ps}}
\end{figure}

Let us take a closer look at the source of $\uonea$ violation. If we write the $\pi$ and $\delta$ correlators (Eqs.~\eqref{eq:pi_delta_corr}) in terms of their left- and right-handed components, we get
\begin{equation}
\begin{split}
C_{\delta/\pi}(x) &=   \big\langle \ubl\dr(x)\dbr\ul(0) + \ubr\dl(x)\dbl\ur(0) \big\rangle\\
                  &\pm \big\langle \ubl\dr(x)\dbl\ur(0) + \ubr\dl(x)\dbr\ul(0) \big\rangle.
\end{split}
\label{eq:c_sps}
\end{equation}
Here the left- and right-handed parts are defined as
\begin{align}
&& \ul(x) = \left(\frac{1-\gamma_5}{2}\right)u(x), && \ur(x) = \left(\frac{1+\gamma_5}{2}\right)u(x), && \notag \\
&& \dl(x) = \left(\frac{1-\gamma_5}{2}\right)d(x), && \dr(x) = \left(\frac{1+\gamma_5}{2}\right)d(x), && 
\end{align}
and
\begin{align}
&& \ubl(x) = \ub(x)\left(\frac{1+\gamma_5}{2}\right), && \ubr(x) = \ub(x)\left(\frac{1-\gamma_5}{2}\right), && \notag \\
&& \dbl(x) = \db(x)\left(\frac{1+\gamma_5}{2}\right), && \dbr(x) = \db(x)\left(\frac{1-\gamma_5}{2}\right). && 
\end{align}
In terms of these, our scalar and pseudoscalar correlators are
\begin{align}
&& \ub(x)d(x) = \ubl(x)\dr(x) + \ubr(x)\dl(x) &&\text{and}&& \ub(x)\gamma_5 d(x) = \ubl(x)\dr(x) - \ubr(x)\dl(x). &&
\end{align}
A $\uonea$ transformation is given by
\begin{align}
&& \ul(x) \to \E^{-\I\theta}\ul(x), && \ubr(x) \to \ubr(x)\E^{-\I\theta}, && \notag \\
&& \ur(x) \to \E^{+\I\theta}\ur(x), && \ubl(x) \to \ubl(x)\E^{+\I\theta}, &&
\end{align}
and similarly for $d(x)$. Looking back at Eq.~\eqref{eq:c_sps}, we see that $\uonea$ violation comes entirely from the terms on the second line, which occur with opposite signs for the correlators. By contrast the terms on the first line, which occur with the same sign for both correlators, are invariant with respect to $\uonea$ transformations. From this it is clear that  the $\uonea$ violating and respecting parts may be isolated by looking at the $\pi-\delta$ and the $\pi+\delta$ correlators respectively.

We plot the sum and the difference of these two correlators in Fig.~\ref{fig:s_pm_ps}. We see that the difference is of the same order of magnitude as the sum at the farthest separations ($x\approx\Ns/2$) for all the temperatures shown here. This reaffirms our earlier observation (Fig.~\ref{fig:susceptibilities}) that $\uonea$ remains broken even at the highest temperatures that we studied.

\begin{figure}[!bth]
\includegraphics[width=0.45\textwidth]{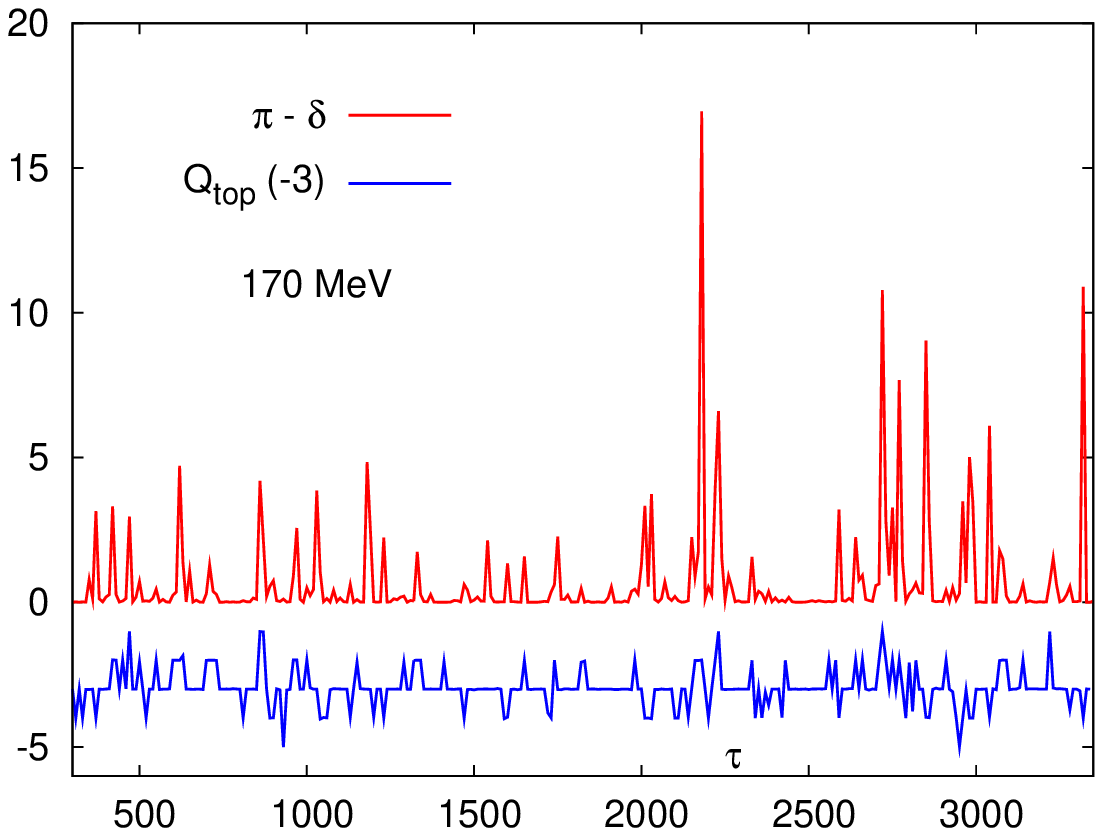}%
\hspace{0.05\textwidth}%
\includegraphics[width=0.45\textwidth]{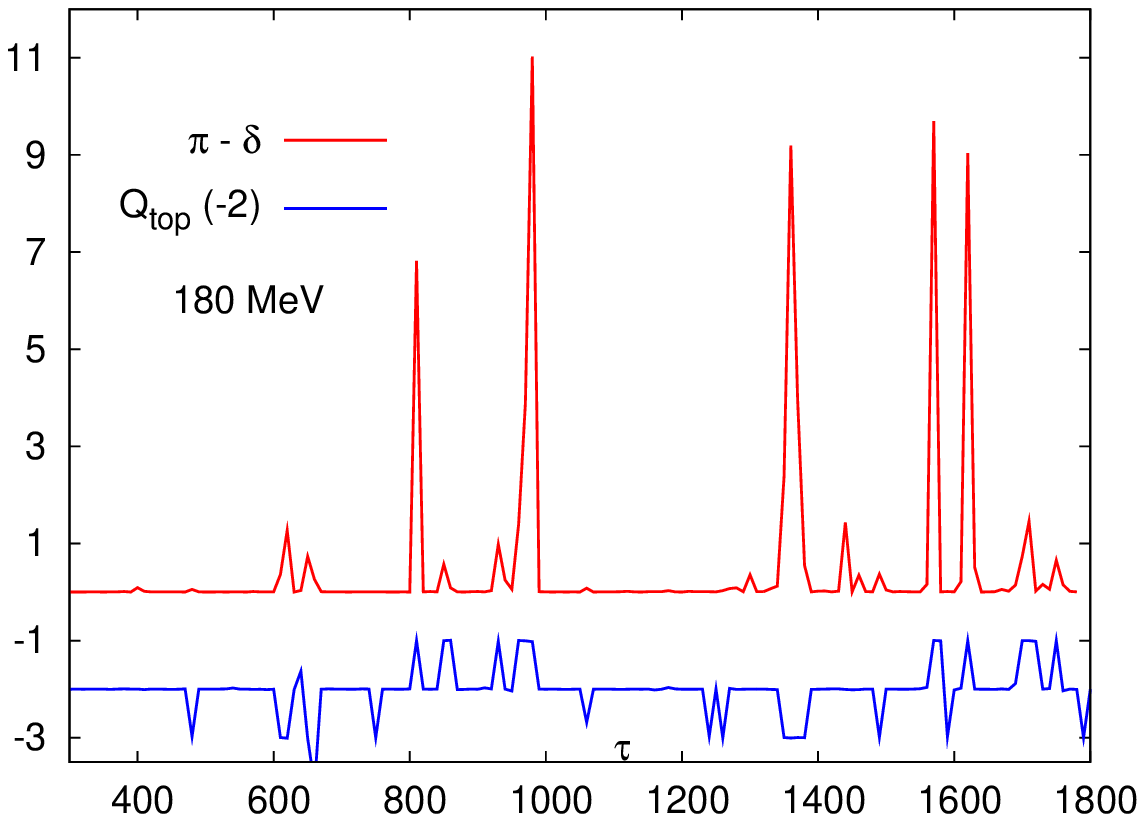}
\includegraphics[width=0.45\textwidth]{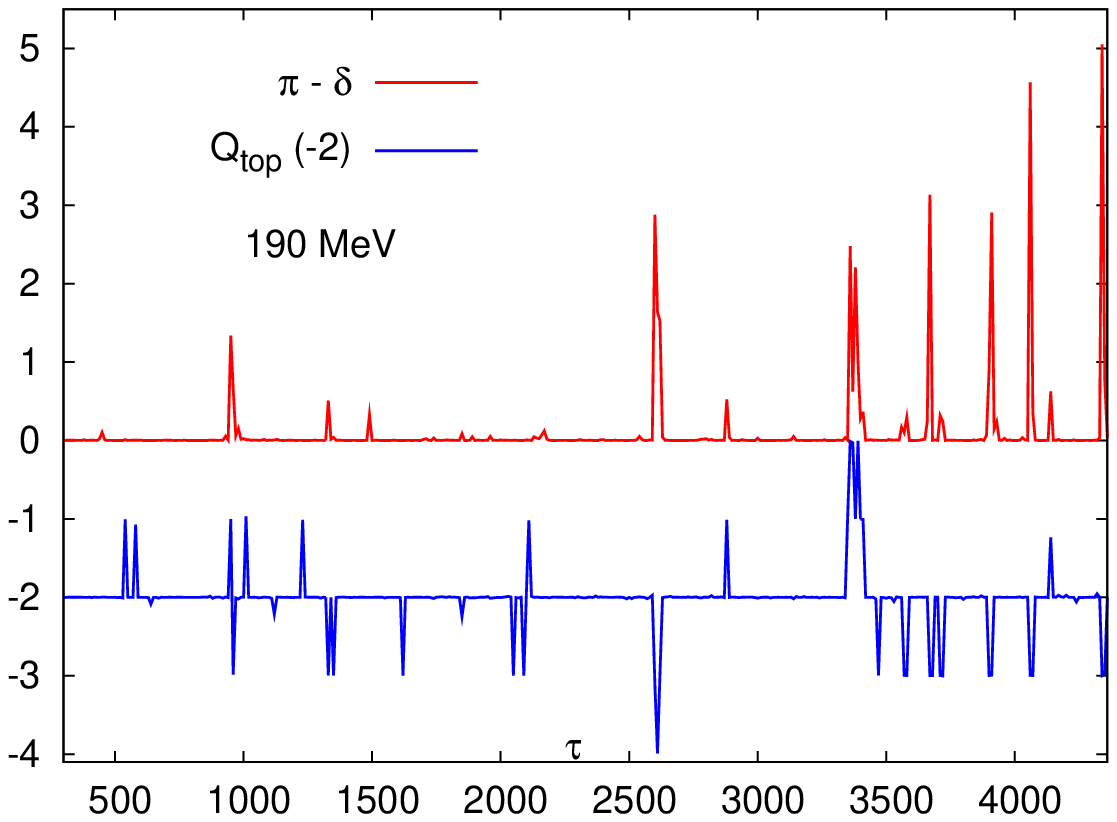}%
\hspace{0.05\textwidth}%
\includegraphics[width=0.45\textwidth]{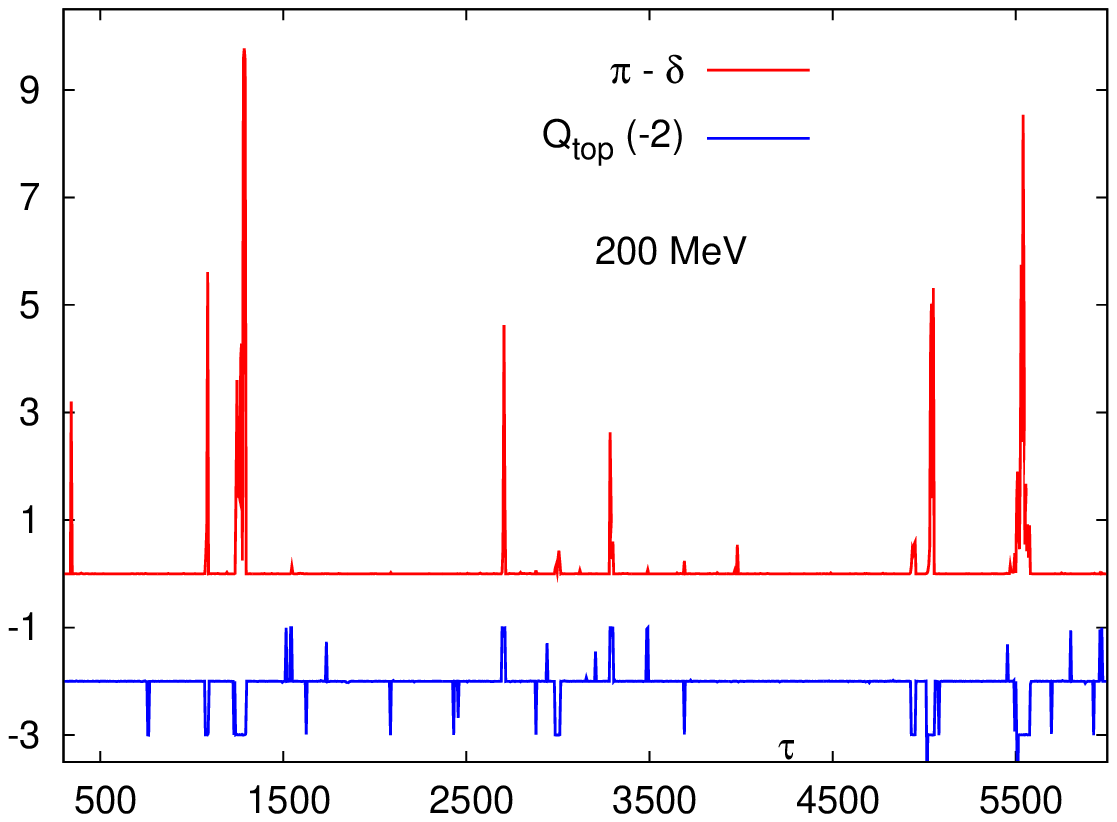}
\caption{Time histories of the integrated correlators (red lines) and the topological charge (blue lines) for $T=170$--200 MeV. The topological charge histories have been displaced downward by 2-3 units for clarity.\label{fig:sps_qtop}}
\end{figure}

It is important to determine whether the $\uonea$ breaking that we observe is due to the presence of topologically nontrivial configurations or occurs merely because of the quark mass. The connection to topology is confirmed when we look at the time histories of these correlators and compare these with the time histories of $\qtop$. Fig.~\ref{fig:sps_qtop} shows these time histories for four temperatures viz. 170, 180, 190 and 200 MeV. To remove the dependence on a particular separation $x$, we plot the time histories of the integrated correlators viz. $\sum_x\left(\pi(x)-\delta(x)\right)\equiv\pi-\delta$. In all but a few cases, the spikes in $\pi-\delta$ are found to line up with the jumps to nonzero values of the topology. In other words, $\uonea$ is broken not ``on average'' but rather by specific configurations whose frequency decreases as the temperature is increased.

\section{The Spectrum of the Dirac Operator}
\label{sec:dirac_spectrum}
The connection to topology is intriguing, but it also raises questions about the eventual fate of $\uonea$ breaking. Nontrivial topologies are distinguished by the fact that the Dirac operator always has a zero eigenvalue in their presence.\footnote{The Atiyah-Singer theorem constrains the \emph{difference} between the number of left- and right-handed zero modes viz. $N_+-N_-=\qtop$. Out of the total $N_++N_-$ zero modes however, only $\qtop$ are stable with respect to small deformations. A stronger statement therefore is that in the presence of a configuration with winding number $\qtop$, the Dirac operator has $\qtop$ \emph{exact} or \emph{robust} zero modes.} The contribution of these modes however vanishes when the four-volume is sent to infinity i.e. in the thermodynamic limit.

To see this, let us express $\lpbpr$ and $\pmd$ in terms of the eigenvalues $\lambda$ of the Dirac operator viz.
\begin{subequations}
\begin{align}
      \lpbpr         &=\int_0^\infty\df\lambda \density\frac{2m}{m^2+\lambda^2}
                      +\frac{\big\langle\lvert\qtop\rvert\big\rangle}{mV},  \label{eq:spectral-pbp}\\
\pmd &=\int_0^\infty\df\lambda \density\frac{4m^2}{\left(m^2+\lambda^2\right)^2}
                      +\frac{2\big\langle\lvert\qtop\rvert\big\rangle}{m^2V}\label{eq:spectral-ua1}.
\end{align}
\label{eq:spectral}
\end{subequations}
The usual weighted average over gauge fields is re-expressed as an average over eigenvalues distributed according to the \emph{spectral density} $\density$. The second term on each RHS represents the contributions of the exact zero modes~\cite{Edwards:1998wx}. The first term represents the contributions coming from the rest of the spectrum.

For large volumes, the topological charge is expected to obey a Gaussian distribution with a width proportional to the volume viz.~\cite{hep-lat/0309189}
\begin{equation}
P(\qtop) = \frac{1}{\sqrt{2\pi\chit V}}\,\exp\left(-\frac{\qtop^2}{2\chit V}\right).
\label{eq:qtop_distr}
\end{equation}
Eq.~\eqref{eq:qtop_distr} implies that $\langle\lvert\qtop\rvert\rangle \propto \sqrt{V}$, hence the second terms in Eq.~\eqref{eq:spectral} vanish as $V\to\infty$. 

$\uonea$-breaking then must come from the rest of the spectrum i.e.~from the first terms of Eqs.~\eqref{eq:spectral}.  In the chiral limit the dominant contribution to these integrals comes from the eigenvalues within a small distance of the origin. This is similar to what happens when chiral symmetry is broken: Eigenvalues $\lambda\sim\mathcal{O}(1/V)$ i.e. the \emph{near-zero} modes, build up near the origin and it is these, rather than the exact zero modes, that break chiral symmetry~\cite{Leutwyler:1992yt}. This is reflected in the Casher-Banks relation $\lpbpr = \pi\rho(0,0)$ for e.g.~\cite{Banks:1979yr}.

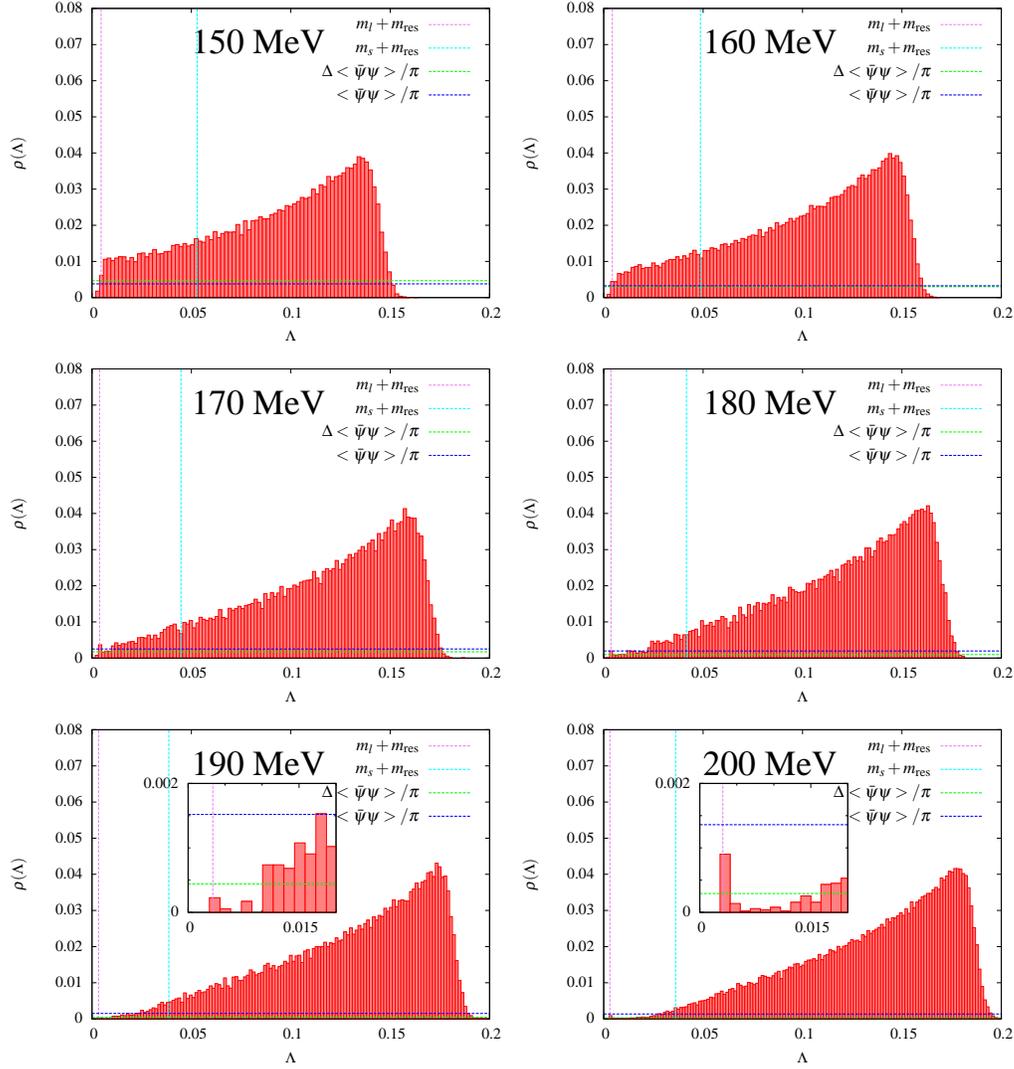
\begin{figure}[htb]
  \centering
  \begin{minipage}[t]{0.45\linewidth}
    \centering
    \resizebox{\linewidth}{!}{\input{./figs/150MeV_1_norm.tex}}
  \end{minipage}%
  \begin{minipage}[t]{0.45\linewidth}
    \centering
    \resizebox{\linewidth}{!}{\input{./figs/160MeV_norm.tex}}
  \end{minipage}
  \begin{minipage}[t]{0.45\linewidth}
    \centering
    \resizebox{\linewidth}{!}{\input{./figs/170MeV_norm.tex}}
  \end{minipage}%
  \begin{minipage}[t]{0.45\linewidth}
    \centering
    \resizebox{\linewidth}{!}{\input{./figs/180MeV_norm.tex}}
  \end{minipage}
  \begin{minipage}[t]{0.45\linewidth}
    \centering
    \resizebox{\linewidth}{!}{\input{./figs/190MeV_norm.tex}}
  \end{minipage}%
  \begin{minipage}[t]{0.45\linewidth}
    \centering
    \resizebox{\linewidth}{!}{\input{./figs/200MeV_1_norm.tex}}
  \end{minipage}
  \caption{(Left to right, top to bottom) The renormalized eigenvalue spectrum for $T=160$ -- 200 MeV. These figures have been taken from~\cite{Lin:2011bj}. The histograms are w.r.t. $\Lambda$, which in the continuum is related to the conventional Dirac eigenvalue $\lambda$ by $\Lambda=\sqrt{\lambda^2+\left(m_l+\mres\right)^2}$. The leftmost line in each plot marks the location of $\Lambda=m_l+\mres$. In the continuum, the eigenvalue density at that point yields the value of $\rho(0,m_l+\mres)$.\label{fig:spectrum}}
\end{figure}

The spectral density $\density$, for small $\lambda$, can be determined by looking at the distribution of the lowest eigenvalues of the Dirac operator with respect to the gauge configurations. For domain wall fermions, the correct Dirac operator is the four-dimensional one whose exact form unfortunately is unknown. However, since it is realized in the low-energy limit of the five-dimensional theory, its low-lying spectrum will be the same as that of the full five-dimensional theory upto an overall renormalization factor which may also be determined non-perturbatively. We show the resulting histograms for $\rho(\lambda)$ in Fig.~\ref{fig:spectrum}. These results, as well as details of the renormalization procedure, have all been presented by Z.~Lin at this conference~\cite{Lin:2011bj} and will also be described in a forthcoming publication~\cite{Bazavov:2012aa}.

From Fig.~\ref{fig:spectrum}, we see that the while the eigenvalue density at the origin shrinks dramatically in going across the chiral phase transition from 150 to 170 MeV, small eigenvalues still occur with reasonable frequency upto $T=180$ MeV. On the other hand, for $T=190$ and 200 MeV most of the eigenvalues occur away from the origin. Nevertheless at both temperatures there is also a second set of eigenvalues that occurs close to the origin. It is these eigenvalues that are responsible for $\uonea$ breaking.

\begin{figure}[!htb]
\centering
\includegraphics[width=0.5\textwidth]{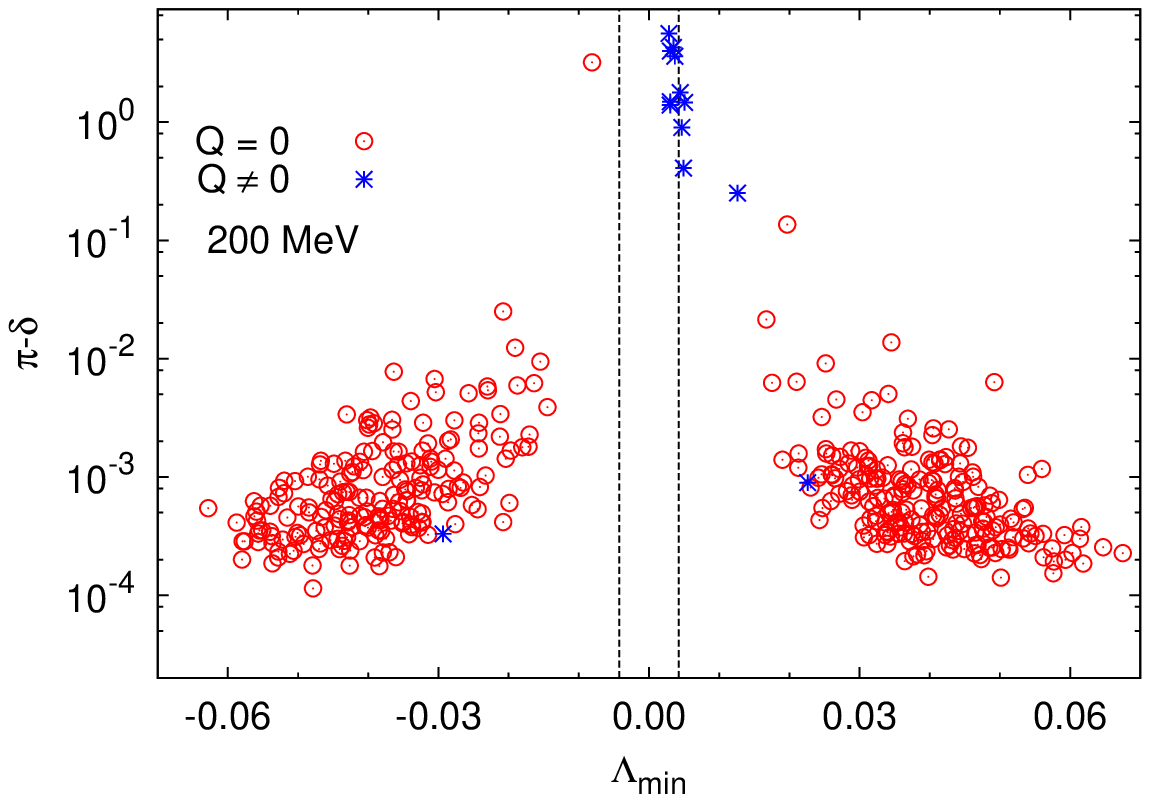}
\caption{Scatter plot showing the correlation between $\pi-\delta$ ($y$-axis) and the value of the smallest eigenvalue $\Lambda_\text{min}$ ($x$-axis) at $T=200$ MeV on a configuration-by-configuration basis. $\Lambda_\text{min}$ is related to the smallest Dirac eigenvalue by $\Lambda_\text{min} = \sqrt{\lmin^2+\left(m_l+\mres\right)^2}$. The dotted vertical lines bracket the unphysical region $\Lambda_\text{min}\leqslant m_l+\mres$. Open circles denote configurations with $\qtop=0$ while bursts denote configurations with $\qtop\neq 0$. Configurations with large values of $\pi-\delta$ are also those with $\qtop\neq 0$, indicating that $\uonea$ breaking at this temperature comes from the exact zero modes. Note that the $y$-axis is logarithmic.}
\label{fig:scatter}
\end{figure}

Fig.~\ref{fig:scatter} shows the correlation between the magnitude of $\uonea$ breaking (i.e. the  value of the integrated correlator $\pi-\delta$) and the value of the smallest eigenvalue for the Dirac operator on a configuration-by-configuration basis. It is clear from the plot that the closer $\lmin$ is to zero, the larger the value of $\pi-\delta$ for that configuration. Furthermore, $\lmin$ is well-separated from zero for the configurations with $\qtop=0$. Accordingly for such configurations $\pi-\delta$ is quite small.

The dominant contribution to $\pmd$ comes from the configurations with $\qtop\neq0$. As expected, $\lmin$ is quite close to zero for these configurations. These eigenvalues however shall vanish in the thermodynamic limit, and the magnitude of $\uonea$ breaking at $T=200$ MeV and 190 MeV is likely to be much smaller than what we currently see.
 
\section{Summary and Discussion}
\label{sec:summary}
The question as to whether $\uonea$ is restored at very high temperatures has a long history. As already mentioned, such a restoration cannot be complete below $T=\infty$. Nevertheless its breaking is very much suppressed at high temperatures and, depending on the magnitude of suppression, it could be \emph{effectively} restored above some temperature. It is natural to ask whether this temperature is close to the familiar chiral transition temperature. In this work, we investigated this question in the context of the $2+1$-flavor theory, on the lattice, by working with a chiral action and examining the behavior of the scalar and pseudoscalar iso-triplet correlators.

We found that $\uonea$ remained broken even after the usual $\sunfa$ chiral symmetry had been restored. Moreover this was due to the presence of configurations with $\qtop\neq0$. Although the proportion of such configurations in our ensembles decreased as the temperature was increased, the difference $\pmd$ was still nonzero at all our temperatures. 

A study of the spectrum of the Dirac operator revealed that this breaking, at least at our highest temperatures ($T=190$ MeV and 200 MeV), resulted from the presence of zero modes which must always arise whenever the underlying gauge configuration has $\qtop\neq0$. Unfortunately since the density of the exact zero modes vanishes as $V^{-1/2}$, where $V$ is the four-volume, the observed breaking is unlikely to persist in the infinite-volume limit.

On the other hand, the density of the \emph{near-zero modes} is an intensive quantity. It is these modes that are responsible for keeping $\uonea$ broken even after chiral symmetry has been restored. The possible form of the spectral density $\density$ that yields $\lpbpr=0$ but $\pmd\neq0$ is an intriguing and currently open question.

\section*{Acknowledgements}
I would like to thank the members of the HotQCD Collaboration, especially Frithjof Karsch, Norman Christ, Swagato Mukherjee, Michael Cheng and Zhongjie Lin, for their help and advice. This work was supported in part by contract DE-AC02-98CH10886 of the U.S. Department of Energy (DoE). The simulations were carried out on the BG/P machine at LLNL, the DOE- and RIKEN-funded QCDOC machines at BNL and the NYBlue machine at the New York Center for Computational Sciences (NYCCS).

\end{document}

%% file: figs/150MeV_1_norm.tex
\begingroup
  \makeatletter
  \providecommand\color[2][]{%
    \GenericError{(gnuplot) \space\space\space\@spaces}{%
      Package color not loaded in conjunction with
      terminal option `colourtext'%
    }{See the gnuplot documentation for explanation.%
    }{Either use 'blacktext' in gnuplot or load the package
      color.sty in LaTeX.}%
    \renewcommand\color[2][]{}%
  }%
  \providecommand\includegraphics[2][]{%
    \GenericError{(gnuplot) \space\space\space\@spaces}{%
      Package graphicx or graphics not loaded%
    }{See the gnuplot documentation for explanation.%
    }{The gnuplot epslatex terminal needs graphicx.sty or graphics.sty.}%
    \renewcommand\includegraphics[2][]{}%
  }%
  \providecommand\rotatebox[2]{#2}%
  \@ifundefined{ifGPcolor}{%
    \newif\ifGPcolor
    \GPcolortrue
  }{}%
  \@ifundefined{ifGPblacktext}{%
    \newif\ifGPblacktext
    \GPblacktexttrue
  }{}%
  \let\gplgaddtomacro\g@addto@macro
  \gdef\gplbacktext{}%
  \gdef\gplfronttext{}%
  \makeatother
  \ifGPblacktext
    \def\colorrgb#1{}%
    \def\colorgray#1{}%
  \else
    \ifGPcolor
      \def\colorrgb#1{\color[rgb]{#1}}%
      \def\colorgray#1{\color[gray]{#1}}%
      \expandafter\def\csname LTw\endcsname{\color{white}}%
      \expandafter\def\csname LTb\endcsname{\color{black}}%
      \expandafter\def\csname LTa\endcsname{\color{black}}%
      \expandafter\def\csname LT0\endcsname{\color[rgb]{1,0,0}}%
      \expandafter\def\csname LT1\endcsname{\color[rgb]{0,1,0}}%
      \expandafter\def\csname LT2\endcsname{\color[rgb]{0,0,1}}%
      \expandafter\def\csname LT3\endcsname{\color[rgb]{1,0,1}}%
      \expandafter\def\csname LT4\endcsname{\color[rgb]{0,1,1}}%
      \expandafter\def\csname LT5\endcsname{\color[rgb]{1,1,0}}%
      \expandafter\def\csname LT6\endcsname{\color[rgb]{0,0,0}}%
      \expandafter\def\csname LT7\endcsname{\color[rgb]{1,0.3,0}}%
      \expandafter\def\csname LT8\endcsname{\color[rgb]{0.5,0.5,0.5}}%
    \else
      \def\colorrgb#1{\color{black}}%
      \def\colorgray#1{\color[gray]{#1}}%
      \expandafter\def\csname LTw\endcsname{\color{white}}%
      \expandafter\def\csname LTb\endcsname{\color{black}}%
      \expandafter\def\csname LTa\endcsname{\color{black}}%
      \expandafter\def\csname LT0\endcsname{\color{black}}%
      \expandafter\def\csname LT1\endcsname{\color{black}}%
      \expandafter\def\csname LT2\endcsname{\color{black}}%
      \expandafter\def\csname LT3\endcsname{\color{black}}%
      \expandafter\def\csname LT4\endcsname{\color{black}}%
      \expandafter\def\csname LT5\endcsname{\color{black}}%
      \expandafter\def\csname LT6\endcsname{\color{black}}%
      \expandafter\def\csname LT7\endcsname{\color{black}}%
      \expandafter\def\csname LT8\endcsname{\color{black}}%
    \fi
  \fi
  \setlength{\unitlength}{0.0500bp}%
  \begin{picture}(7200.00,5040.00)%
    \gplgaddtomacro\gplbacktext{%
      \csname LTb\endcsname%
      \put(1078,704){\makebox(0,0)[r]{\strut{} 0}}%
      \put(1078,1213){\makebox(0,0)[r]{\strut{} 0.01}}%
      \put(1078,1722){\makebox(0,0)[r]{\strut{} 0.02}}%
      \put(1078,2231){\makebox(0,0)[r]{\strut{} 0.03}}%
      \put(1078,2740){\makebox(0,0)[r]{\strut{} 0.04}}%
      \put(1078,3248){\makebox(0,0)[r]{\strut{} 0.05}}%
      \put(1078,3757){\makebox(0,0)[r]{\strut{} 0.06}}%
      \put(1078,4266){\makebox(0,0)[r]{\strut{} 0.07}}%
      \put(1078,4775){\makebox(0,0)[r]{\strut{} 0.08}}%
      \put(1210,484){\makebox(0,0){\strut{} 0}}%
      \put(2608,484){\makebox(0,0){\strut{} 0.05}}%
      \put(4007,484){\makebox(0,0){\strut{} 0.1}}%
      \put(5405,484){\makebox(0,0){\strut{} 0.15}}%
      \put(6803,484){\makebox(0,0){\strut{} 0.2}}%
      \put(176,2739){\rotatebox{-270}{\makebox(0,0){\strut{}$\rho(\Lambda)$}}}%
      \put(4006,154){\makebox(0,0){\strut{}$\Lambda$}}%
      \put(2608,4266){\makebox(0,0)[l]{\strut{}\Huge 150 MeV}}%
    }%
    \gplgaddtomacro\gplfronttext{%
      \csname LTb\endcsname%
      \put(5816,4547){\makebox(0,0)[r]{\strut{}\large $ m_l+m_{ \mathrm{res} } $}}%
      \csname LTb\endcsname%
      \put(5816,4217){\makebox(0,0)[r]{\strut{}\large $ m_s+m_{ \mathrm{res} } $}}%
      \csname LTb\endcsname%
      \put(5816,3887){\makebox(0,0)[r]{\strut{}\large $\Delta<\bar{\psi}\psi>/\pi$}}%
      \csname LTb\endcsname%
      \put(5816,3557){\makebox(0,0)[r]{\strut{}\large $<\bar{\psi}\psi>/\pi$}}%
    }%
    \gplbacktext
    \put(0,0){\includegraphics{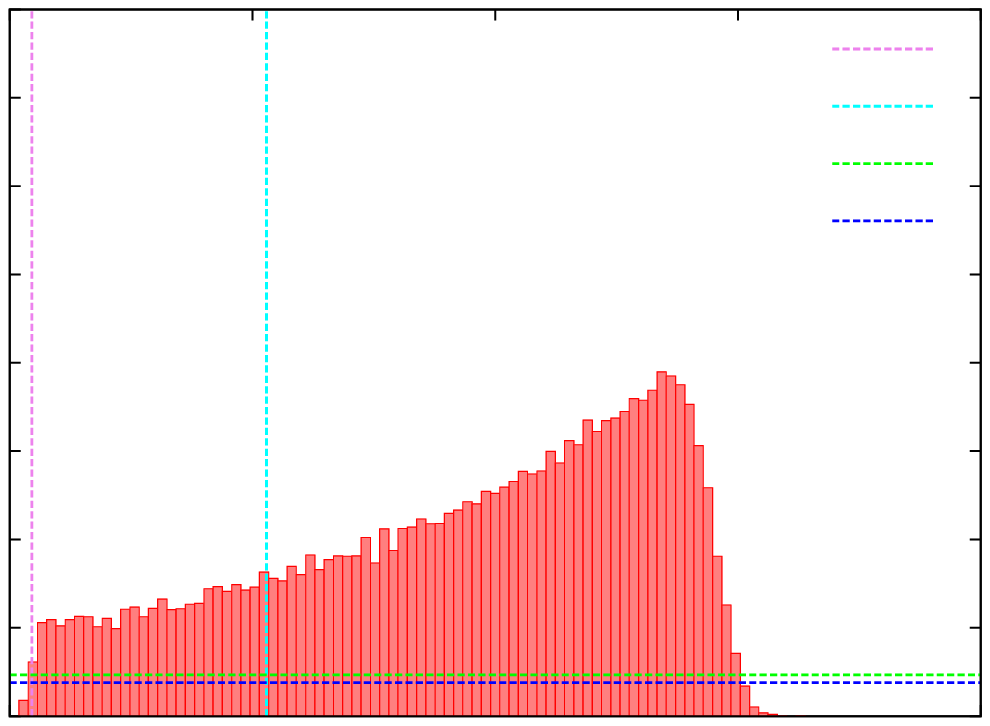}}%
    \gplfronttext
  \end{picture}%
\endgroup

%% file: figs/160MeV_norm.tex
\begingroup
  \makeatletter
  \providecommand\color[2][]{%
    \GenericError{(gnuplot) \space\space\space\@spaces}{%
      Package color not loaded in conjunction with
      terminal option `colourtext'%
    }{See the gnuplot documentation for explanation.%
    }{Either use 'blacktext' in gnuplot or load the package
      color.sty in LaTeX.}%
    \renewcommand\color[2][]{}%
  }%
  \providecommand\includegraphics[2][]{%
    \GenericError{(gnuplot) \space\space\space\@spaces}{%
      Package graphicx or graphics not loaded%
    }{See the gnuplot documentation for explanation.%
    }{The gnuplot epslatex terminal needs graphicx.sty or graphics.sty.}%
    \renewcommand\includegraphics[2][]{}%
  }%
  \providecommand\rotatebox[2]{#2}%
  \@ifundefined{ifGPcolor}{%
    \newif\ifGPcolor
    \GPcolortrue
  }{}%
  \@ifundefined{ifGPblacktext}{%
    \newif\ifGPblacktext
    \GPblacktexttrue
  }{}%
  \let\gplgaddtomacro\g@addto@macro
  \gdef\gplbacktext{}%
  \gdef\gplfronttext{}%
  \makeatother
  \ifGPblacktext
    \def\colorrgb#1{}%
    \def\colorgray#1{}%
  \else
    \ifGPcolor
      \def\colorrgb#1{\color[rgb]{#1}}%
      \def\colorgray#1{\color[gray]{#1}}%
      \expandafter\def\csname LTw\endcsname{\color{white}}%
      \expandafter\def\csname LTb\endcsname{\color{black}}%
      \expandafter\def\csname LTa\endcsname{\color{black}}%
      \expandafter\def\csname LT0\endcsname{\color[rgb]{1,0,0}}%
      \expandafter\def\csname LT1\endcsname{\color[rgb]{0,1,0}}%
      \expandafter\def\csname LT2\endcsname{\color[rgb]{0,0,1}}%
      \expandafter\def\csname LT3\endcsname{\color[rgb]{1,0,1}}%
      \expandafter\def\csname LT4\endcsname{\color[rgb]{0,1,1}}%
      \expandafter\def\csname LT5\endcsname{\color[rgb]{1,1,0}}%
      \expandafter\def\csname LT6\endcsname{\color[rgb]{0,0,0}}%
      \expandafter\def\csname LT7\endcsname{\color[rgb]{1,0.3,0}}%
      \expandafter\def\csname LT8\endcsname{\color[rgb]{0.5,0.5,0.5}}%
    \else
      \def\colorrgb#1{\color{black}}%
      \def\colorgray#1{\color[gray]{#1}}%
      \expandafter\def\csname LTw\endcsname{\color{white}}%
      \expandafter\def\csname LTb\endcsname{\color{black}}%
      \expandafter\def\csname LTa\endcsname{\color{black}}%
      \expandafter\def\csname LT0\endcsname{\color{black}}%
      \expandafter\def\csname LT1\endcsname{\color{black}}%
      \expandafter\def\csname LT2\endcsname{\color{black}}%
      \expandafter\def\csname LT3\endcsname{\color{black}}%
      \expandafter\def\csname LT4\endcsname{\color{black}}%
      \expandafter\def\csname LT5\endcsname{\color{black}}%
      \expandafter\def\csname LT6\endcsname{\color{black}}%
      \expandafter\def\csname LT7\endcsname{\color{black}}%
      \expandafter\def\csname LT8\endcsname{\color{black}}%
    \fi
  \fi
  \setlength{\unitlength}{0.0500bp}%
  \begin{picture}(7200.00,5040.00)%
    \gplgaddtomacro\gplbacktext{%
      \csname LTb\endcsname%
      \put(1078,704){\makebox(0,0)[r]{\strut{} 0}}%
      \put(1078,1213){\makebox(0,0)[r]{\strut{} 0.01}}%
      \put(1078,1722){\makebox(0,0)[r]{\strut{} 0.02}}%
      \put(1078,2231){\makebox(0,0)[r]{\strut{} 0.03}}%
      \put(1078,2740){\makebox(0,0)[r]{\strut{} 0.04}}%
      \put(1078,3248){\makebox(0,0)[r]{\strut{} 0.05}}%
      \put(1078,3757){\makebox(0,0)[r]{\strut{} 0.06}}%
      \put(1078,4266){\makebox(0,0)[r]{\strut{} 0.07}}%
      \put(1078,4775){\makebox(0,0)[r]{\strut{} 0.08}}%
      \put(1210,484){\makebox(0,0){\strut{} 0}}%
      \put(2608,484){\makebox(0,0){\strut{} 0.05}}%
      \put(4007,484){\makebox(0,0){\strut{} 0.1}}%
      \put(5405,484){\makebox(0,0){\strut{} 0.15}}%
      \put(6803,484){\makebox(0,0){\strut{} 0.2}}%
      \put(176,2739){\rotatebox{-270}{\makebox(0,0){\strut{}$\rho(\Lambda)$}}}%
      \put(4006,154){\makebox(0,0){\strut{}$\Lambda$}}%
      \put(2608,4266){\makebox(0,0)[l]{\strut{}\Huge 160 MeV}}%
    }%
    \gplgaddtomacro\gplfronttext{%
      \csname LTb\endcsname%
      \put(5816,4547){\makebox(0,0)[r]{\strut{}\large $ m_l+m_{ \mathrm{res} } $}}%
      \csname LTb\endcsname%
      \put(5816,4217){\makebox(0,0)[r]{\strut{}\large $ m_s+m_{ \mathrm{res} } $}}%
      \csname LTb\endcsname%
      \put(5816,3887){\makebox(0,0)[r]{\strut{}\large $\Delta<\bar{\psi}\psi>/\pi$}}%
      \csname LTb\endcsname%
      \put(5816,3557){\makebox(0,0)[r]{\strut{}\large $<\bar{\psi}\psi>/\pi$}}%
    }%
    \gplbacktext
    \put(0,0){\includegraphics{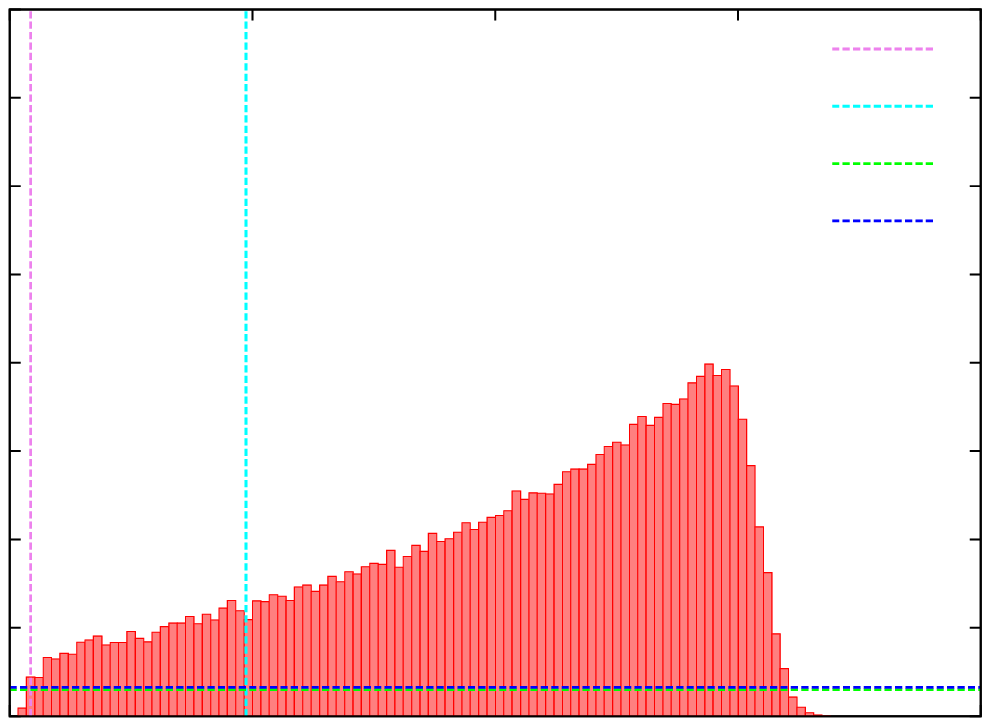}}%
    \gplfronttext
  \end{picture}%
\endgroup

%% file: figs/170MeV_norm.tex
\begingroup
  \makeatletter
  \providecommand\color[2][]{%
    \GenericError{(gnuplot) \space\space\space\@spaces}{%
      Package color not loaded in conjunction with
      terminal option `colourtext'%
    }{See the gnuplot documentation for explanation.%
    }{Either use 'blacktext' in gnuplot or load the package
      color.sty in LaTeX.}%
    \renewcommand\color[2][]{}%
  }%
  \providecommand\includegraphics[2][]{%
    \GenericError{(gnuplot) \space\space\space\@spaces}{%
      Package graphicx or graphics not loaded%
    }{See the gnuplot documentation for explanation.%
    }{The gnuplot epslatex terminal needs graphicx.sty or graphics.sty.}%
    \renewcommand\includegraphics[2][]{}%
  }%
  \providecommand\rotatebox[2]{#2}%
  \@ifundefined{ifGPcolor}{%
    \newif\ifGPcolor
    \GPcolortrue
  }{}%
  \@ifundefined{ifGPblacktext}{%
    \newif\ifGPblacktext
    \GPblacktexttrue
  }{}%
  \let\gplgaddtomacro\g@addto@macro
  \gdef\gplbacktext{}%
  \gdef\gplfronttext{}%
  \makeatother
  \ifGPblacktext
    \def\colorrgb#1{}%
    \def\colorgray#1{}%
  \else
    \ifGPcolor
      \def\colorrgb#1{\color[rgb]{#1}}%
      \def\colorgray#1{\color[gray]{#1}}%
      \expandafter\def\csname LTw\endcsname{\color{white}}%
      \expandafter\def\csname LTb\endcsname{\color{black}}%
      \expandafter\def\csname LTa\endcsname{\color{black}}%
      \expandafter\def\csname LT0\endcsname{\color[rgb]{1,0,0}}%
      \expandafter\def\csname LT1\endcsname{\color[rgb]{0,1,0}}%
      \expandafter\def\csname LT2\endcsname{\color[rgb]{0,0,1}}%
      \expandafter\def\csname LT3\endcsname{\color[rgb]{1,0,1}}%
      \expandafter\def\csname LT4\endcsname{\color[rgb]{0,1,1}}%
      \expandafter\def\csname LT5\endcsname{\color[rgb]{1,1,0}}%
      \expandafter\def\csname LT6\endcsname{\color[rgb]{0,0,0}}%
      \expandafter\def\csname LT7\endcsname{\color[rgb]{1,0.3,0}}%
      \expandafter\def\csname LT8\endcsname{\color[rgb]{0.5,0.5,0.5}}%
    \else
      \def\colorrgb#1{\color{black}}%
      \def\colorgray#1{\color[gray]{#1}}%
      \expandafter\def\csname LTw\endcsname{\color{white}}%
      \expandafter\def\csname LTb\endcsname{\color{black}}%
      \expandafter\def\csname LTa\endcsname{\color{black}}%
      \expandafter\def\csname LT0\endcsname{\color{black}}%
      \expandafter\def\csname LT1\endcsname{\color{black}}%
      \expandafter\def\csname LT2\endcsname{\color{black}}%
      \expandafter\def\csname LT3\endcsname{\color{black}}%
      \expandafter\def\csname LT4\endcsname{\color{black}}%
      \expandafter\def\csname LT5\endcsname{\color{black}}%
      \expandafter\def\csname LT6\endcsname{\color{black}}%
      \expandafter\def\csname LT7\endcsname{\color{black}}%
      \expandafter\def\csname LT8\endcsname{\color{black}}%
    \fi
  \fi
  \setlength{\unitlength}{0.0500bp}%
  \begin{picture}(7200.00,5040.00)%
    \gplgaddtomacro\gplbacktext{%
      \csname LTb\endcsname%
      \put(1078,704){\makebox(0,0)[r]{\strut{} 0}}%
      \put(1078,1213){\makebox(0,0)[r]{\strut{} 0.01}}%
      \put(1078,1722){\makebox(0,0)[r]{\strut{} 0.02}}%
      \put(1078,2231){\makebox(0,0)[r]{\strut{} 0.03}}%
      \put(1078,2740){\makebox(0,0)[r]{\strut{} 0.04}}%
      \put(1078,3248){\makebox(0,0)[r]{\strut{} 0.05}}%
      \put(1078,3757){\makebox(0,0)[r]{\strut{} 0.06}}%
      \put(1078,4266){\makebox(0,0)[r]{\strut{} 0.07}}%
      \put(1078,4775){\makebox(0,0)[r]{\strut{} 0.08}}%
      \put(1210,484){\makebox(0,0){\strut{} 0}}%
      \put(2608,484){\makebox(0,0){\strut{} 0.05}}%
      \put(4007,484){\makebox(0,0){\strut{} 0.1}}%
      \put(5405,484){\makebox(0,0){\strut{} 0.15}}%
      \put(6803,484){\makebox(0,0){\strut{} 0.2}}%
      \put(176,2739){\rotatebox{-270}{\makebox(0,0){\strut{}$\rho(\Lambda)$}}}%
      \put(4006,154){\makebox(0,0){\strut{}$\Lambda$}}%
      \put(2608,4266){\makebox(0,0)[l]{\strut{}\Huge 170 MeV}}%
    }%
    \gplgaddtomacro\gplfronttext{%
      \csname LTb\endcsname%
      \put(5816,4547){\makebox(0,0)[r]{\strut{}\large $ m_l+m_{ \mathrm{res} } $}}%
      \csname LTb\endcsname%
      \put(5816,4217){\makebox(0,0)[r]{\strut{}\large $ m_s+m_{ \mathrm{res} } $}}%
      \csname LTb\endcsname%
      \put(5816,3887){\makebox(0,0)[r]{\strut{}\large $\Delta<\bar{\psi}\psi>/\pi$}}%
      \csname LTb\endcsname%
      \put(5816,3557){\makebox(0,0)[r]{\strut{}\large $<\bar{\psi}\psi>/\pi$}}%
    }%
    \gplbacktext
    \put(0,0){\includegraphics{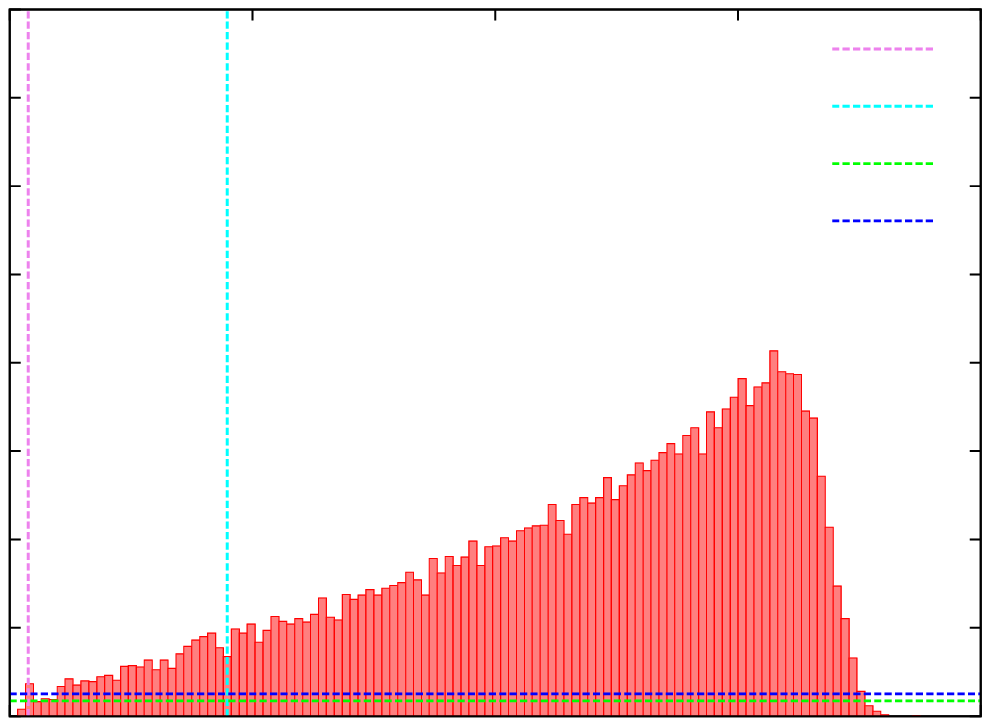}}%
    \gplfronttext
  \end{picture}%
\endgroup

%% file: figs/180MeV_norm.tex
\begingroup
  \makeatletter
  \providecommand\color[2][]{%
    \GenericError{(gnuplot) \space\space\space\@spaces}{%
      Package color not loaded in conjunction with
      terminal option `colourtext'%
    }{See the gnuplot documentation for explanation.%
    }{Either use 'blacktext' in gnuplot or load the package
      color.sty in LaTeX.}%
    \renewcommand\color[2][]{}%
  }%
  \providecommand\includegraphics[2][]{%
    \GenericError{(gnuplot) \space\space\space\@spaces}{%
      Package graphicx or graphics not loaded%
    }{See the gnuplot documentation for explanation.%
    }{The gnuplot epslatex terminal needs graphicx.sty or graphics.sty.}%
    \renewcommand\includegraphics[2][]{}%
  }%
  \providecommand\rotatebox[2]{#2}%
  \@ifundefined{ifGPcolor}{%
    \newif\ifGPcolor
    \GPcolortrue
  }{}%
  \@ifundefined{ifGPblacktext}{%
    \newif\ifGPblacktext
    \GPblacktexttrue
  }{}%
  \let\gplgaddtomacro\g@addto@macro
  \gdef\gplbacktext{}%
  \gdef\gplfronttext{}%
  \makeatother
  \ifGPblacktext
    \def\colorrgb#1{}%
    \def\colorgray#1{}%
  \else
    \ifGPcolor
      \def\colorrgb#1{\color[rgb]{#1}}%
      \def\colorgray#1{\color[gray]{#1}}%
      \expandafter\def\csname LTw\endcsname{\color{white}}%
      \expandafter\def\csname LTb\endcsname{\color{black}}%
      \expandafter\def\csname LTa\endcsname{\color{black}}%
      \expandafter\def\csname LT0\endcsname{\color[rgb]{1,0,0}}%
      \expandafter\def\csname LT1\endcsname{\color[rgb]{0,1,0}}%
      \expandafter\def\csname LT2\endcsname{\color[rgb]{0,0,1}}%
      \expandafter\def\csname LT3\endcsname{\color[rgb]{1,0,1}}%
      \expandafter\def\csname LT4\endcsname{\color[rgb]{0,1,1}}%
      \expandafter\def\csname LT5\endcsname{\color[rgb]{1,1,0}}%
      \expandafter\def\csname LT6\endcsname{\color[rgb]{0,0,0}}%
      \expandafter\def\csname LT7\endcsname{\color[rgb]{1,0.3,0}}%
      \expandafter\def\csname LT8\endcsname{\color[rgb]{0.5,0.5,0.5}}%
    \else
      \def\colorrgb#1{\color{black}}%
      \def\colorgray#1{\color[gray]{#1}}%
      \expandafter\def\csname LTw\endcsname{\color{white}}%
      \expandafter\def\csname LTb\endcsname{\color{black}}%
      \expandafter\def\csname LTa\endcsname{\color{black}}%
      \expandafter\def\csname LT0\endcsname{\color{black}}%
      \expandafter\def\csname LT1\endcsname{\color{black}}%
      \expandafter\def\csname LT2\endcsname{\color{black}}%
      \expandafter\def\csname LT3\endcsname{\color{black}}%
      \expandafter\def\csname LT4\endcsname{\color{black}}%
      \expandafter\def\csname LT5\endcsname{\color{black}}%
      \expandafter\def\csname LT6\endcsname{\color{black}}%
      \expandafter\def\csname LT7\endcsname{\color{black}}%
      \expandafter\def\csname LT8\endcsname{\color{black}}%
    \fi
  \fi
  \setlength{\unitlength}{0.0500bp}%
  \begin{picture}(7200.00,5040.00)%
    \gplgaddtomacro\gplbacktext{%
      \csname LTb\endcsname%
      \put(1078,704){\makebox(0,0)[r]{\strut{} 0}}%
      \put(1078,1213){\makebox(0,0)[r]{\strut{} 0.01}}%
      \put(1078,1722){\makebox(0,0)[r]{\strut{} 0.02}}%
      \put(1078,2231){\makebox(0,0)[r]{\strut{} 0.03}}%
      \put(1078,2740){\makebox(0,0)[r]{\strut{} 0.04}}%
      \put(1078,3248){\makebox(0,0)[r]{\strut{} 0.05}}%
      \put(1078,3757){\makebox(0,0)[r]{\strut{} 0.06}}%
      \put(1078,4266){\makebox(0,0)[r]{\strut{} 0.07}}%
      \put(1078,4775){\makebox(0,0)[r]{\strut{} 0.08}}%
      \put(1210,484){\makebox(0,0){\strut{} 0}}%
      \put(2608,484){\makebox(0,0){\strut{} 0.05}}%
      \put(4007,484){\makebox(0,0){\strut{} 0.1}}%
      \put(5405,484){\makebox(0,0){\strut{} 0.15}}%
      \put(6803,484){\makebox(0,0){\strut{} 0.2}}%
      \put(176,2739){\rotatebox{-270}{\makebox(0,0){\strut{}$\rho(\Lambda)$}}}%
      \put(4006,154){\makebox(0,0){\strut{}$\Lambda$}}%
      \put(2608,4266){\makebox(0,0)[l]{\strut{}\Huge 180 MeV}}%
    }%
    \gplgaddtomacro\gplfronttext{%
      \csname LTb\endcsname%
      \put(5816,4547){\makebox(0,0)[r]{\strut{}\large $ m_l+m_{ \mathrm{res} } $}}%
      \csname LTb\endcsname%
      \put(5816,4217){\makebox(0,0)[r]{\strut{}\large $ m_s+m_{ \mathrm{res} } $}}%
      \csname LTb\endcsname%
      \put(5816,3887){\makebox(0,0)[r]{\strut{}\large $\Delta<\bar{\psi}\psi>/\pi$}}%
      \csname LTb\endcsname%
      \put(5816,3557){\makebox(0,0)[r]{\strut{}\large $<\bar{\psi}\psi>/\pi$}}%
    }%
    \gplbacktext
    \put(0,0){\includegraphics{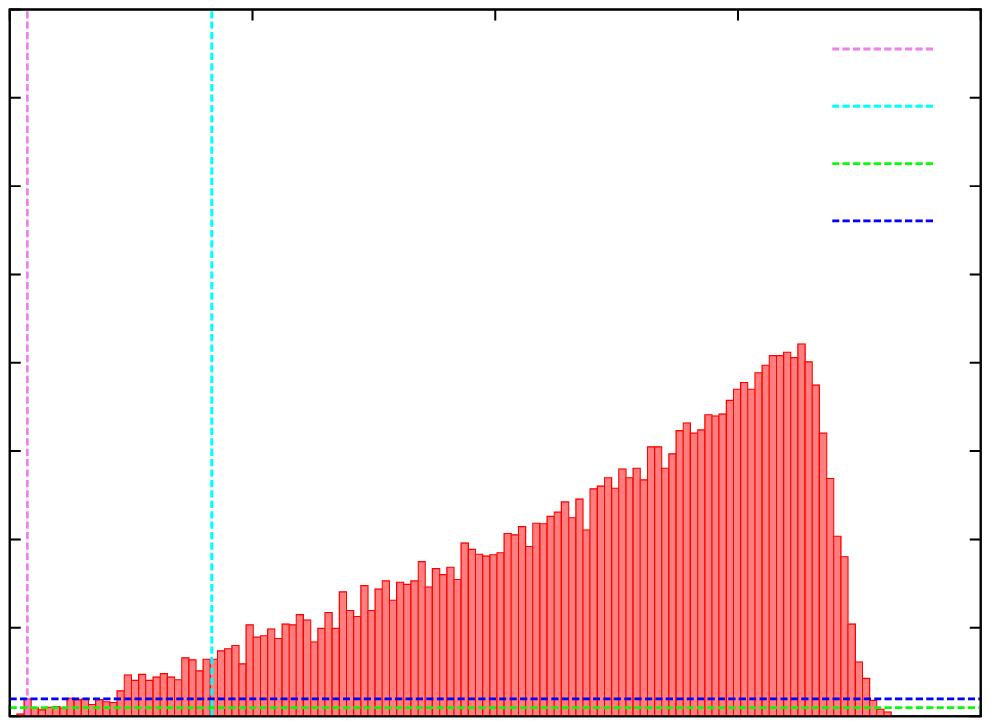}}%
    \gplfronttext
  \end{picture}%
\endgroup

%% file: figs/190MeV_norm.tex
\begingroup
  \makeatletter
  \providecommand\color[2][]{%
    \GenericError{(gnuplot) \space\space\space\@spaces}{%
      Package color not loaded in conjunction with
      terminal option `colourtext'%
    }{See the gnuplot documentation for explanation.%
    }{Either use 'blacktext' in gnuplot or load the package
      color.sty in LaTeX.}%
    \renewcommand\color[2][]{}%
  }%
  \providecommand\includegraphics[2][]{%
    \GenericError{(gnuplot) \space\space\space\@spaces}{%
      Package graphicx or graphics not loaded%
    }{See the gnuplot documentation for explanation.%
    }{The gnuplot epslatex terminal needs graphicx.sty or graphics.sty.}%
    \renewcommand\includegraphics[2][]{}%
  }%
  \providecommand\rotatebox[2]{#2}%
  \@ifundefined{ifGPcolor}{%
    \newif\ifGPcolor
    \GPcolortrue
  }{}%
  \@ifundefined{ifGPblacktext}{%
    \newif\ifGPblacktext
    \GPblacktexttrue
  }{}%
  \let\gplgaddtomacro\g@addto@macro
  \gdef\gplbacktext{}%
  \gdef\gplfronttext{}%
  \makeatother
  \ifGPblacktext
    \def\colorrgb#1{}%
    \def\colorgray#1{}%
  \else
    \ifGPcolor
      \def\colorrgb#1{\color[rgb]{#1}}%
      \def\colorgray#1{\color[gray]{#1}}%
      \expandafter\def\csname LTw\endcsname{\color{white}}%
      \expandafter\def\csname LTb\endcsname{\color{black}}%
      \expandafter\def\csname LTa\endcsname{\color{black}}%
      \expandafter\def\csname LT0\endcsname{\color[rgb]{1,0,0}}%
      \expandafter\def\csname LT1\endcsname{\color[rgb]{0,1,0}}%
      \expandafter\def\csname LT2\endcsname{\color[rgb]{0,0,1}}%
      \expandafter\def\csname LT3\endcsname{\color[rgb]{1,0,1}}%
      \expandafter\def\csname LT4\endcsname{\color[rgb]{0,1,1}}%
      \expandafter\def\csname LT5\endcsname{\color[rgb]{1,1,0}}%
      \expandafter\def\csname LT6\endcsname{\color[rgb]{0,0,0}}%
      \expandafter\def\csname LT7\endcsname{\color[rgb]{1,0.3,0}}%
      \expandafter\def\csname LT8\endcsname{\color[rgb]{0.5,0.5,0.5}}%
    \else
      \def\colorrgb#1{\color{black}}%
      \def\colorgray#1{\color[gray]{#1}}%
      \expandafter\def\csname LTw\endcsname{\color{white}}%
      \expandafter\def\csname LTb\endcsname{\color{black}}%
      \expandafter\def\csname LTa\endcsname{\color{black}}%
      \expandafter\def\csname LT0\endcsname{\color{black}}%
      \expandafter\def\csname LT1\endcsname{\color{black}}%
      \expandafter\def\csname LT2\endcsname{\color{black}}%
      \expandafter\def\csname LT3\endcsname{\color{black}}%
      \expandafter\def\csname LT4\endcsname{\color{black}}%
      \expandafter\def\csname LT5\endcsname{\color{black}}%
      \expandafter\def\csname LT6\endcsname{\color{black}}%
      \expandafter\def\csname LT7\endcsname{\color{black}}%
      \expandafter\def\csname LT8\endcsname{\color{black}}%
    \fi
  \fi
  \setlength{\unitlength}{0.0500bp}%
  \begin{picture}(7200.00,5040.00)%
    \gplgaddtomacro\gplbacktext{%
      \csname LTb\endcsname%
      \put(1078,704){\makebox(0,0)[r]{\strut{} 0}}%
      \put(1078,1213){\makebox(0,0)[r]{\strut{} 0.01}}%
      \put(1078,1722){\makebox(0,0)[r]{\strut{} 0.02}}%
      \put(1078,2231){\makebox(0,0)[r]{\strut{} 0.03}}%
      \put(1078,2740){\makebox(0,0)[r]{\strut{} 0.04}}%
      \put(1078,3248){\makebox(0,0)[r]{\strut{} 0.05}}%
      \put(1078,3757){\makebox(0,0)[r]{\strut{} 0.06}}%
      \put(1078,4266){\makebox(0,0)[r]{\strut{} 0.07}}%
      \put(1078,4775){\makebox(0,0)[r]{\strut{} 0.08}}%
      \put(1210,484){\makebox(0,0){\strut{} 0}}%
      \put(2608,484){\makebox(0,0){\strut{} 0.05}}%
      \put(4007,484){\makebox(0,0){\strut{} 0.1}}%
      \put(5405,484){\makebox(0,0){\strut{} 0.15}}%
      \put(6803,484){\makebox(0,0){\strut{} 0.2}}%
      \put(176,2739){\rotatebox{-270}{\makebox(0,0){\strut{}$\rho(\Lambda)$}}}%
      \put(4006,154){\makebox(0,0){\strut{}$\Lambda$}}%
      \put(2608,4266){\makebox(0,0)[l]{\strut{}\Huge 190 MeV}}%
    }%
    \gplgaddtomacro\gplfronttext{%
      \csname LTb\endcsname%
      \put(5816,4547){\makebox(0,0)[r]{\strut{}\large $ m_l+m_{ \mathrm{res} } $}}%
      \csname LTb\endcsname%
      \put(5816,4217){\makebox(0,0)[r]{\strut{}\large $ m_s+m_{ \mathrm{res} } $}}%
      \csname LTb\endcsname%
      \put(5816,3887){\makebox(0,0)[r]{\strut{}\large $\Delta<\bar{\psi}\psi>/\pi$}}%
      \csname LTb\endcsname%
      \put(5816,3557){\makebox(0,0)[r]{\strut{}\large $<\bar{\psi}\psi>/\pi$}}%
    }%
    \gplgaddtomacro\gplbacktext{%
      \csname LTb\endcsname%
      \put(2430,2204){\makebox(0,0)[r]{\strut{} 0}}%
      \put(2430,4019){\makebox(0,0)[r]{\strut{} 0.002}}%
      \put(2562,1984){\makebox(0,0){\strut{} 0}}%
      \put(4123,1984){\makebox(0,0){\strut{} 0.015}}%
      \put(7765,65729){\makebox(0,0)[l]{\strut{}\Huge IamMeV}}%
    }%
    \gplgaddtomacro\gplfronttext{%
    }%
    \gplbacktext
    \put(0,0){\includegraphics{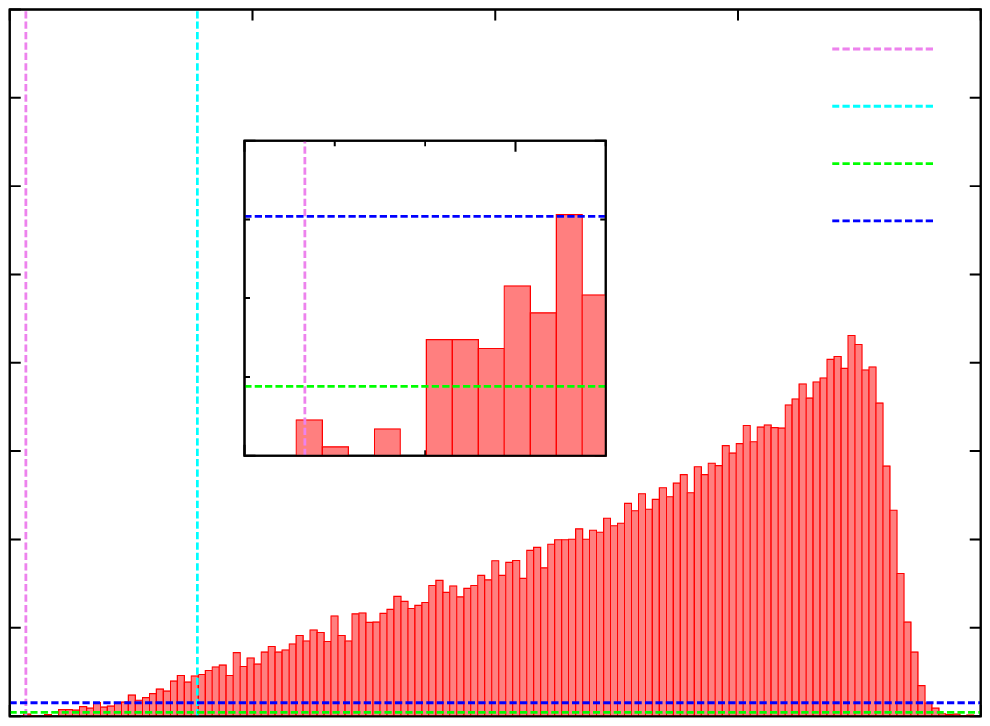}}%
    \gplfronttext
  \end{picture}%
\endgroup

%% file: figs/200MeV_1_norm.tex
\begingroup
  \makeatletter
  \providecommand\color[2][]{%
    \GenericError{(gnuplot) \space\space\space\@spaces}{%
      Package color not loaded in conjunction with
      terminal option `colourtext'%
    }{See the gnuplot documentation for explanation.%
    }{Either use 'blacktext' in gnuplot or load the package
      color.sty in LaTeX.}%
    \renewcommand\color[2][]{}%
  }%
  \providecommand\includegraphics[2][]{%
    \GenericError{(gnuplot) \space\space\space\@spaces}{%
      Package graphicx or graphics not loaded%
    }{See the gnuplot documentation for explanation.%
    }{The gnuplot epslatex terminal needs graphicx.sty or graphics.sty.}%
    \renewcommand\includegraphics[2][]{}%
  }%
  \providecommand\rotatebox[2]{#2}%
  \@ifundefined{ifGPcolor}{%
    \newif\ifGPcolor
    \GPcolortrue
  }{}%
  \@ifundefined{ifGPblacktext}{%
    \newif\ifGPblacktext
    \GPblacktexttrue
  }{}%
  \let\gplgaddtomacro\g@addto@macro
  \gdef\gplbacktext{}%
  \gdef\gplfronttext{}%
  \makeatother
  \ifGPblacktext
    \def\colorrgb#1{}%
    \def\colorgray#1{}%
  \else
    \ifGPcolor
      \def\colorrgb#1{\color[rgb]{#1}}%
      \def\colorgray#1{\color[gray]{#1}}%
      \expandafter\def\csname LTw\endcsname{\color{white}}%
      \expandafter\def\csname LTb\endcsname{\color{black}}%
      \expandafter\def\csname LTa\endcsname{\color{black}}%
      \expandafter\def\csname LT0\endcsname{\color[rgb]{1,0,0}}%
      \expandafter\def\csname LT1\endcsname{\color[rgb]{0,1,0}}%
      \expandafter\def\csname LT2\endcsname{\color[rgb]{0,0,1}}%
      \expandafter\def\csname LT3\endcsname{\color[rgb]{1,0,1}}%
      \expandafter\def\csname LT4\endcsname{\color[rgb]{0,1,1}}%
      \expandafter\def\csname LT5\endcsname{\color[rgb]{1,1,0}}%
      \expandafter\def\csname LT6\endcsname{\color[rgb]{0,0,0}}%
      \expandafter\def\csname LT7\endcsname{\color[rgb]{1,0.3,0}}%
      \expandafter\def\csname LT8\endcsname{\color[rgb]{0.5,0.5,0.5}}%
    \else
      \def\colorrgb#1{\color{black}}%
      \def\colorgray#1{\color[gray]{#1}}%
      \expandafter\def\csname LTw\endcsname{\color{white}}%
      \expandafter\def\csname LTb\endcsname{\color{black}}%
      \expandafter\def\csname LTa\endcsname{\color{black}}%
      \expandafter\def\csname LT0\endcsname{\color{black}}%
      \expandafter\def\csname LT1\endcsname{\color{black}}%
      \expandafter\def\csname LT2\endcsname{\color{black}}%
      \expandafter\def\csname LT3\endcsname{\color{black}}%
      \expandafter\def\csname LT4\endcsname{\color{black}}%
      \expandafter\def\csname LT5\endcsname{\color{black}}%
      \expandafter\def\csname LT6\endcsname{\color{black}}%
      \expandafter\def\csname LT7\endcsname{\color{black}}%
      \expandafter\def\csname LT8\endcsname{\color{black}}%
    \fi
  \fi
  \setlength{\unitlength}{0.0500bp}%
  \begin{picture}(7200.00,5040.00)%
    \gplgaddtomacro\gplbacktext{%
      \csname LTb\endcsname%
      \put(1078,704){\makebox(0,0)[r]{\strut{} 0}}%
      \put(1078,1213){\makebox(0,0)[r]{\strut{} 0.01}}%
      \put(1078,1722){\makebox(0,0)[r]{\strut{} 0.02}}%
      \put(1078,2231){\makebox(0,0)[r]{\strut{} 0.03}}%
      \put(1078,2740){\makebox(0,0)[r]{\strut{} 0.04}}%
      \put(1078,3248){\makebox(0,0)[r]{\strut{} 0.05}}%
      \put(1078,3757){\makebox(0,0)[r]{\strut{} 0.06}}%
      \put(1078,4266){\makebox(0,0)[r]{\strut{} 0.07}}%
      \put(1078,4775){\makebox(0,0)[r]{\strut{} 0.08}}%
      \put(1210,484){\makebox(0,0){\strut{} 0}}%
      \put(2608,484){\makebox(0,0){\strut{} 0.05}}%
      \put(4007,484){\makebox(0,0){\strut{} 0.1}}%
      \put(5405,484){\makebox(0,0){\strut{} 0.15}}%
      \put(6803,484){\makebox(0,0){\strut{} 0.2}}%
      \put(176,2739){\rotatebox{-270}{\makebox(0,0){\strut{}$\rho(\Lambda)$}}}%
      \put(4006,154){\makebox(0,0){\strut{}$\Lambda$}}%
      \put(2608,4266){\makebox(0,0)[l]{\strut{}\Huge 200 MeV}}%
    }%
    \gplgaddtomacro\gplfronttext{%
      \csname LTb\endcsname%
      \put(5816,4547){\makebox(0,0)[r]{\strut{}\large $ m_l+m_{ \mathrm{res} } $}}%
      \csname LTb\endcsname%
      \put(5816,4217){\makebox(0,0)[r]{\strut{}\large $ m_s+m_{ \mathrm{res} } $}}%
      \csname LTb\endcsname%
      \put(5816,3887){\makebox(0,0)[r]{\strut{}\large $\Delta<\bar{\psi}\psi>/\pi$}}%
      \csname LTb\endcsname%
      \put(5816,3557){\makebox(0,0)[r]{\strut{}\large $<\bar{\psi}\psi>/\pi$}}%
    }%
    \gplgaddtomacro\gplbacktext{%
      \csname LTb\endcsname%
      \put(2430,2204){\makebox(0,0)[r]{\strut{} 0}}%
      \put(2430,4019){\makebox(0,0)[r]{\strut{} 0.002}}%
      \put(2562,1984){\makebox(0,0){\strut{} 0}}%
      \put(4123,1984){\makebox(0,0){\strut{} 0.015}}%
      \put(7765,65729){\makebox(0,0)[l]{\strut{}\Huge IamMeV}}%
    }%
    \gplgaddtomacro\gplfronttext{%
    }%
    \gplbacktext
    \put(0,0){\includegraphics{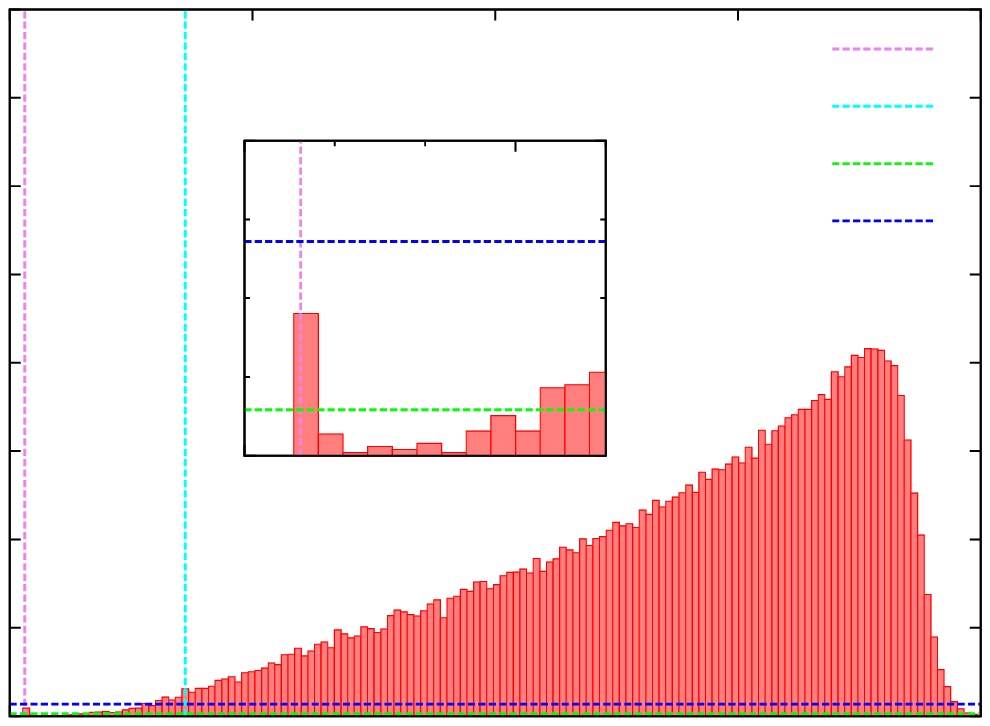}}%
    \gplfronttext
  \end{picture}%
\endgroup